\documentclass[copyright,creativecommons]{eptcs}
\providecommand{\event}{VPT 2016} 
\newif\iflncs\lncsfalse
\newif\ifconf\conftrue
\usepackage[leqno]{amsmath}
\usepackage{amsthm}
\usepackage{amssymb}
\usepackage{wasysym}
\usepackage{paralist}
\usepackage{fancyvrb}
\iflncs\else\usepackage{url}\fi
\usepackage{stmaryrd}
\usepackage[draft]{fixme}
\usepackage{xspace}
\usepackage{wrapfig}
\usepackage{graphicx}
\usepackage[all]{xy}
\usepackage{tikz}
\usepackage{tkz-graph}
\usetikzlibrary{arrows,automata,positioning,fit}
\usetikzlibrary{calc}
\usetikzlibrary{shapes,decorations}
\usetikzlibrary{decorations.pathmorphing}
\usepackage{listings}
\iflncs
\else
\ifconf
\else
\fi\fi

\iflncs
\else
\newtheorem{theorem}{THEOREM}[section]
\newtheorem{lemma}[theorem]{LEMMA}
\newtheorem{obs}[theorem]{Observation}

\newtheorem{corol}[theorem]{Corollary}
\theoremstyle{definition}
\newtheorem{definition}[theorem]{Definition}

\newtheorem{example}[theorem]{EXAMPLE}
\fi

\newcommand{\pgt}[1]{{\tt #1}}
\newcommand{\semi}{\texttt{;}}
\newcommand{\pass}{\mathrel{\mbox{\texttt{:=}}}}
\newcommand{\wpass}{\mathrel{:\le}}
\newcommand{\lsem}{\mbox{$\lbrack\hspace{-0.3ex}\lbrack$}}
\newcommand{\rsem}{\mbox{$\rbrack\hspace{-0.3ex}\rbrack$}}

\newcommand{\C}{{\mbox{\upshape\texttt{C}}}\xspace}
\newcommand{\X}{{\mbox{\upshape\texttt{X}}}}
\newcommand{\Y}{{\mbox{\upshape\texttt{Y}}}}

\let\com=\newcommand
\com{\be}{\begin{enumerate}}
\com{\ee}{\end{enumerate}}
\com{\bi}{\begin{itemize}}
\com{\ei}{\end{itemize}}
\com{\bthm}{\begin{theorem}}
\com{\ethm}{\end{theorem}}
\com{\bdfn}{\begin{definition}}
\com{\edfn}{\end{definition}}
\com{\blem}{\begin{lemma}}
\com{\elem}{\end{lemma}}
\com{\bex}{\begin{example}}
\com{\eex}{\end{example}}
\newcommand{\bprf}{\begin{proof}}
\newcommand{\eprf}{\end{proof}}
\newcommand{\bobs}{\begin{obs}}
\newcommand{\eobs}{\end{obs}}
\newcommand{\bcor}{\begin{corol}}
\newcommand{\ecor}{\end{corol}}

\newenvironment{alg}
{\par\medskip\par\noindent\textbf{Algorithm}\ }{\par\noindent}

\newenvironment{fig0}[3]
{
\xdef\fighack{\noexpand\caption{#1}\noexpand\label{#3}}
\begin{figure}[#2]
\hspace*{3mm}\begin{minipage}{0.95\columnwidth}}{\end{minipage}%
\fighack%
\end{figure}}

\newcommand\Lbjk{\ensuremath{L_{\text{BJK}}}\xspace}

\newcommand{\nats}{\mathbb{N}}
\newcommand{\inds}{{\mathbb{I}}}
\newcommand{\prog}{{\mathtt{p}}}
\newcommand{\states}[1]{\mathtt{St}_{#1}}
\newcommand{\locations}[1]{\mathtt{Loc}_{#1}}
\newcommand{\fctransitions}[1]{\mathcal{T}_{#1}}
\newcommand{\fctseqs}[1]{\mathcal{T}^{+}_{#1}}
\newcommand{\fctseqsbnd}[1]{\mathcal{T}^{\oplus}_{#1}}
\newcommand{\arcs}[1]{\mathtt{Arc}_{#1}}
\newcommand{\bound}[1]{\mathtt{Bound}(#1)}
\newcommand{\cutset}[1]{\mathtt{Cutset}({#1})}
\newcommand{\instr}[1]{\mathtt{Inst}({#1})}
\newcommand{\instset}{\Sigma}
\newcommand{\loops}[1]{{\mathcal L}_{#1}}

\newcommand{\trans}[3]{{#2\xrightarrow{#1}#3}}
\newcommand{\tseq}[3]{{#1\stackrel{#2}{\leadsto}#3}}

\newcommand\st{\colon} 
\newcommand{\spacedfive}[5]{#1\quad #2\quad #3\quad #4\quad #5}   

\newcommand{\tseqs}{{\mathcal{T}}}
\newcommand{\deptypes}{{\mathbb{D}}}

\newcommand{\unarydep}[3]{{#1\xrightarrow{\makebox[1ex][l]{$\scriptstyle #2$}} #3}}
\newcommand{\longunarydep}[3]{{#1\xrightarrow{#2} #3}}
\newcommand{\binarydep}[4]{{\textstyle {{#1}\atop{#2}}\rightrightarrows {{#3}\atop{#4}}}}
\newcommand{\depsdice}{{\mathbb{F}}}
\newcommand\matrices{{{\deptypes_0}^{(n+1)\times (n+1)}}}

\com{\eqdef}{\stackrel{\mathrm{def}}{=}}

\newcommand{\laregen}{\mbox{\textrm{LARE}}\xspace}  
\ifconf
\newcommand{\lare}{\mbox{\textrm{LARE}}\xspace}
\else
\newcommand{\lare}{\mbox{\textrm{LARE}$_0$}\xspace}
\fi

\newcommand{\lareoneE}{\mbox{\textrm{LARE}$_1$}\xspace}
\newcommand{\lareprog}{{\mathtt{e}}}
\newcommand{\ensuresymbol}{{\ensuremath{\cent}}}
\newcommand{\tssempar}[1]{\mbox{\lsem{$#1$}\rsem}_{ts}}
\newcommand{\depsempar}[1]{\mathchoice{\lbrack\hspace{-0.3ex}\lbrack #1\rbrack\hspace{-0.3ex}\rbrack_{dep}}%
{\lbrack\hspace{-0.3ex}\lbrack #1 \rbrack\hspace{-0.3ex}\rbrack_{dep}}%
{\lbrack\hspace{-0.25ex}\lbrack  {\scriptstyle #1} \rbrack\hspace{-0.25ex}\rbrack_{dep}}%
{\lbrack\hspace{-0.25ex}\lbrack  {\scriptscriptstyle #1} \rbrack\hspace{-0.25ex}\rbrack_{dep}}}

\let\labsem=\lsem
\newcommand{\rabsem}{\mbox{$\rbrack\hspace{-0.3ex}\rbrack_{dep}$}}
\newcommand{\absempol}[1]{\mathchoice{\mbox{\labsem $#1$\rabsem}}{\mbox{\labsem $#1$\rabsem}}%
{\lbrack\hspace{-0.25ex}\lbrack  {#1} \rbrack\hspace{-0.25ex}\rbrack_{dep}}%
{\lbrack\hspace{-0.25ex}\lbrack  {#1} \rbrack\hspace{-0.25ex}\rbrack_{dep}}}
\newcommand{\addbdeps}{{\textsc{Complete}}}
\newcommand{\iddep}{{\ensuremath{I_{dep}}}}
\newcommand{\concat}{\fatsemi}   
\newcommand{\erase}[1]{{\textsc{Erase}(#1)}}   
\newcommand{\ecount}[1]{{\Vert #1 \Vert}}
\newcommand{\droploc}[0]{{\textsc{AnonLoc}}}
\newcommand{\addloc}[0]{{\textsc{AddLoc}}}
\newcommand{\subst}[3]{{\textsc{Subst}(#1,#2,#3)}}

\newcommand\xmin{{x_{\it min}}}
\newcommand\ymin{{y_{\it min}}}
\newcommand\zmin{{z_{\it min}}}

\newcommand{\MM}{{\mathcal{M}}}

\newcommand\somsym{{\mathfrak M}}
\newcommand\somsempol[1]{\somsym \absempol{#1} }

\newcommand\spge[1]{\sqsupseteq}  

\newcommand{\LoopExp}{(\ensuresymbol E)^*}

\newcommand{\gbox}[1]{\ensuremath{\,\fbox{#1}\,}}
\newcommand{\askip}[0]{\gbox{\texttt{skip}}}
\newcommand{\acopy}[2]{\gbox{$\X_{#1}\pass\X_{#2}$}}
\newcommand{\asum}[3]{\gbox{$\X_{#1}\pass\X_{#2}+\X_{#3}$}}
\newcommand{\amul}[3]{\gbox{$\X_{#1}\pass\X_{#2}*\X_{#3}$}}

\newcommand{\koat}{\pgt{KoAT}\xspace}

\begin{document}
\ifconf
\def\authorrunning{Ben-Amram, Pineles}
\def\titlerunning{Flowchart Programs}
\fi
\title{Flowchart Programs, Regular Expressions, and Decidability of Polynomial Growth-Rate}
\author{
   Amir M. Ben-Amram
\ifconf\institute{
School of Computer Science, Tel-Aviv Academic College,  Israel}
\email{amirben@mta.ac.il}
\else%
\thanks{School of Computer Science, Tel-Aviv Academic College,  Israel. 
\texttt{amirben@mta.ac.il}}
\fi
   \and
   Aviad Pineles
   \ifconf
   \institute{}
   \email{paviad2@gmail.com}
\fi
}

\maketitle   


\vspace{-20pt}

\begin{abstract}
We present a new method for inferring complexity properties
for a class of programs in the form of flowcharts annotated with loop information.
Specifically, our method can (soundly and completely) decide
if computed values are polynomially bounded as a function of the input; and similarly for the running time.
Such complexity properties are undecidable for
a Turing-complete programming language, and a common work-around 
in program analysis is to settle for sound but incomplete solutions. 
In contrast, 
we consider a class of programs 
that is Turing-incomplete, but strong enough to include
several challenges for this kind of analysis. For a related language that has \emph{well-structured} syntax,
similar to Meyer and Ritchie's LOOP programs,
the problem has been previously proved to be decidable.
The analysis relied on the compositionality of programs, hence the challenge
in obtaining similar results for flowchart programs with arbitrary control-flow graphs. 
Our answer to the challenge is twofold: first, we propose a class of loop-annotated flowcharts, which is more general
than the class of flowcharts that directly represent structured programs;
secondly, we present a technique to reuse the ideas from the work on structured programs and apply them to such flowcharts.
The technique is inspired by the classic translation of non-deterministic automata to regular expressions,
 but we obviate the exponential cost of constructing such an expression, obtaining a polynomial-time analysis.
These ideas may well be applicable to other analysis problems.

\medskip
\flushleft
{\bf Keywords}: polynomial time complexity, cost analysis, program transformation
\end{abstract}

\section{Introduction}

Devising algorithms that deduce complexity properties of a given program is 
a classic problem of program analysis~\cite{Wegbreit:75,Rosendahl89,ACE},
and has received considerable attention in recent years.
 Ideally, a static-analysis tool will
warn us in compilation-time whenever our program fails a complexity specification. 
For instance, we could be warned of algorithms whose running time is not polynomially bounded, or algorithms
that compute super-polynomially large values.

While practical tools typically attempt to infer explicit bounds, as a theoretical problem it is sufficiently 
challenging to look at the classification problem---polynomial or not.
Since deciding such a property for all programs
in a Turing-complete language is impossible, there are two ways to approach the problem. There is much work
(as the above-cited) which targets a Turing-complete language and settles for an incomplete solution. Other works
investigate the decidability of the problem in restricted languages. 
In~\cite{BJK08}, the problem of \emph{polynomial boundedness} is shown decidable 
(moreover, in PTIME) for
a ``core'' imperative language $\Lbjk$ with a bounded loop command%
\footnote{Since we consider explicitly-bounded loops, the problem of classifying the time complexity of the program
is equivalent to the problem of classifying growth rates of variables.}.	
The language of~\cite{BJK08} only has
numeric variables, with the basic operations of addition and multiplication.
Figure~\ref{fig-syntax}
shows the syntax of the language; regarding its semantics,
it is important that the language has a non-deterministic choice instead of 
conditionals, and bounded loops (``do at most $n$ times'') which are also non-deterministic (for further
details see~\cite{BJK08}). 

The design of this core language is influenced by the approach ``abstract and conquer.''
This means that while we are solving a subproblem, this could be used also as a component in a partial solution to the ``real"
problem (handling a realistic language) by using it as a \emph{back end}, assuming that some 
\emph{front end} analyser translates source programs, in a conservative fashion, into core programs.
Abstraction (e.g., making branches non-deterministic by simply hiding the conditionals) is clearly doable,
while sophisticated \emph{static analysis}
technology may allow the construction of more precise front ends.
Importantly, current static analysis techniques allow for computing \emph{loop bounds} 
\cite{Albert-et-al:TCS:2011,ADFG:2010,SPEED-POPL09}
and thus transforming an arbitrary loop into our bounded-loop form.
These intentions motivate the choice of including non-determinism in our core language.

Another reason for making the core language non-deterministic is the desire to find a decidable problem. 
In fact, including precise test as conditionals (say, testing for equality of variables) is easily seen
to lead to undecidability, and \cite{BK11} shows that having a deterministic loop (with precise
iteration count, rather than just a bound) breaks decidability, too.

To carry this research forward, we ask: in what ways can the core language be extended, while maintaining decidability?
In this paper we consider an extension that involves the control structure of the program: we extend from nested loop
programs to \emph{unstructured} (``flowchart") programs. The details of this extension are motivated, as we will
explain, by looking at the way certain complexity analysers for realistic programs work, trying to break the process into
a front-end and a back-end, and make the back-end decision problem solvable. Theoretically, the result is a polynomial-time 
decidability result for a language $FC$ that strictly extends $\Lbjk$.
Our analysis algorithm for the $FC$ language is obtained by a technique that is more general than the particular
application. The technique arose from trying to adapt the analysis of~\cite{BJK08}. The challenge was
that the analysis was compositional 
and essentially based on the well-nested structure of the core language.
How can such an analysis be performed on arbitrary control-flow graphs?
Our solution is guided by the standard transformation of a finite automaton (NFA) to a regular expression.
A regular expression is a sort of well-structured program. In order to make this as explicit as possible, we define
a programming language which is written in a syntax similar to regexp's. However, semantically it is an extension of
$\Lbjk$ (thus, as a by-product, we find a structured language with precisely the expressivity of $FC$). We show that by
composing the NFA-to-regexp transformation with an analysis of the structured language (a natural extension
of the analysis of~\cite{BJK08}) we obtain an analyzer for $FC$ that runs in polynomial time (somewhat surprising,
as the general construction of regular expressions is exponential-time). We believe that many static analysis algorithms 
that process general control-flow graphs may implicitly use the same approach, but that we are first to make it explicit. 

The rest of this paper is structured as follows: first we motivate, then formally define, our concept of loop-annotated flowcharts.
Then, we define the regexp-like language, called Loop-Annotated Regular Expressions (LARE, for short), and explain how we translate flowcharts to
LARE. Finally we give the algorithm that analyzes LARE. Some of the more technical definitions and proofs, and in particular, the 
correctness proofs of the analysis algorithm, which are long and complex, are deferred to appendices.

\begin{figure}[t]
$$ \renewcommand{\arraystretch}{1.3}
\begin{array}{rcl}
\verb+X+,\verb+Y+,\verb+Z+\in\mbox{Variable} &\;\; ::= \;\; & \X_1 \mid\X_2 \mid \X_3 \mid
 \ldots  \mid \X_n\\
\verb+e+\in\mbox{Expression} & ::= & \verb+X+ \mid \verb/Y + Z/ \mid 
\verb+Y * Z+\\
\verb+C+\in\mbox{Command} & ::= & \verb+skip+ \mid \verb+X:=e+ 
                                \mid \verb+C+_1 \semi \verb+C+_2 
                                \mid \texttt{loop} \; \X  \; \texttt{\{C\}} 
                   \mid \texttt{choose}\;  \texttt{C}_1  \; \texttt{or} \;
                   \texttt{C}_2
 \end{array} \renewcommand{\arraystretch}{1.0}$$

\caption{Syntax of the structured core language $\Lbjk$ of [6].}
\label{fig-syntax}
\end{figure}

\section{Loop-Annotated Flowchart Programs: Motivation}
\label{sec:FCmotivations}

%
The algorithm presented in this paper analyzes programs in a language which we call \emph{Loop-Annotated Flowchart Programs}. This language has the
important features that 
\begin{inparaenum}[(i)]
\item program form is an arbitrary control-flow graph with instructions on arcs;
\item variables hold non-negative integers;
\item the instruction set is highly limited; 
\item information about loop bounds is supplied as ``annotations," presented in the next section.
\end{inparaenum}
The design results from two goals, on one hand we
 are looking for a decidable case; on the other hand we are motivated 
by looking at tools that analyze real-world programs. We next explain this motivation informally,
focusing on the tool \cite{ADFG:2010}. The tool generates from the input program  a control-flow graph, where arcs carry both guards and updates;
for example,  the C program shown in Figure~\ref{fig:Cprog} (a)
 might be represented as 
in Figure~\ref{fig:Cprog} (b).

 The tool uses a linear-programming 
based algorithm to find \emph{ranking functions} for parts  (subgraphs) of the control-flow graph. A ranking function is a combination
of the program variables that is non-increasing throughout the subgraph while on certain arcs it is non-negative and strictly decreasing.
This implies
a bound on the number  transitions (program steps) that correspond to the latter kind of arcs.
 If there remains a strongly-connected subgraph for which an iteration bound is not implied,
a ranking function for the subgraph is necessary. For example, in Figure~\ref{fig:Cprog}, the algorithm of~\cite{ADFG:2010}
reports the function $\verb/i/+\verb/n/$, holding throughout the strongly-connected component of the graph, and strictly decreasing
on the bottom arc from (3) to (2); this implies an upper bound of $2\verb/n/$ (the initial value of this
function) on the number of times one can take this arc while staying within the component. The algorithm next excludes this arc and
finds the ranking function $\verb/i/+\verb/j/$  
for the remaining cycle (note that $\verb/j/$ would have sufficed; ranking functions are not unique).
The second function implies the bound $\verb/i/+\verb/n/$ on
the number of iterations through that cycle (in terms of the values of
variables on entrance to this loop). 
Thus, the analysis decomposes the program into two nested loops (Figure~\ref{fig:Cprog} (c)), also establishing, for each loop, an iteration bound
in terms of variables that do not change in the loop.
Note that this decomposition is not evident from the program text (where there is just one loop construct), and is not unique, in general. Moreover,
it depends on semantic analysis and cannot be determined just from the graph structure. 
\emph{Our focus in this work is on the analysis of a program that is already decomposed and annotated with bounds}. 
Thus the above discussion should be understood purely as motivation; we do not deal with the art of abstracting source programs
in realistic languages. Our decidability results only concern questions about a loop-annotated flowcharts. We note that this does not translate
to a decidability result regarding the source programs, since the process of identifying and annotating loops in not well defined (there may be
many correct annotations for  given problem); moreover, undecidable problems may be encoded into this phase.
For simplicity, in our input language, loop bounds will always be specified as
a single variable. This is not a restriction, since if we are considering a given
program where a loop bound is an expression in program variables
(e.g., a sum, as above) we can generate an auxiliary variable to store the loop bound.

Another motivation to handle flowcharts is the usage of abstraction refinement techniques, where the control-flow graph
representing a program is expanded to represent additional properties of the current or past states%
~\cite{rival2007trace}.
 Even when starting from a structured program, the structure of the resulting control-flow graph might not correspond to that of the
source program.
As an example, Pineles~\cite{Pineles:2014} extends the language treated here with reset assignments \verb/X := 0/.
In~\cite{B2010:DICE}, such assignments were added to \Lbjk, which required a non-trivial addition to the
analysis algorithm.
 In contradiction, \cite{Pineles:2014} handles the extension by program transformation:
we refine the control-flow graph with respect to the history of resets.
This allows for eliminating the resets, so the
growth-rate analysis does not have to handle them. However, it also changes the graph structure so it no longer corresponds
to the original, structured program.

\begin{figure}
\flushleft
\begin{minipage}{0.25\textwidth}
\flushleft
\begin{lstlisting}[numbers=none,belowskip=0pt]
assume (n>=0);
i = n;
j = n;
while (i > 0) {
  if (j>0) {
    j = j-1;
  } else {
    j = n;
    i = i-1;
  }
}
\end{lstlisting}
\end{minipage}
\tikzset{every state/.style={minimum size=1.8em},align at top/.style={baseline=(current bounding box.north)}}%
\begin{minipage}{0.7\textwidth}
\raisebox{-\height}{\begin{tikzpicture}[
  -triangle 45,
  auto,
  node distance=2cm,
  thick,
  scale=0.6
]
  \node[state] (A)                     {$1$};
  \node[state] (B) [right of=A]        {$2$};
  \node[state] (C) [above of=B]  {$4$};
  \node[state] (D)  at ($(B)+(4.5cm,0cm)$)  {$3$};
  
  \path[every node/.style={anchor=south,auto=false,align=center}]
        (A) edge node {\lstinline/i=n; j=n/} (B)
        (B) edge node [anchor=east] {\lstinline/[i<=0]/} (C)
        (B) edge node [anchor=south] {\raisebox{-2pt}[2pt][0pt]{\lstinline/[i>0]/}} (D)
        (D) edge[bend right=30]  node (backarc1) [anchor=south] {\lstinline/[j>0]; j = j-1/} (B)
        (D) edge[bend left=30]  node (backarc2) [anchor=north] {\lstinline/j = n; i = i-1/} (B);
\end{tikzpicture}}
\hspace{1cm}
\raisebox{-\height}{\begin{tikzpicture}[
  -triangle 45,
  auto,
  node distance=2cm,
  thick,
  scale=0.6
]
  \node[state] (A)                     {$1$};
  \node[state] (B) [right of=A]        {$2$};
  \node[state] (C) [above of=B]  {$4$};
  \node[state] (D)  at ($(B)+(3.5cm,0cm)$)  {$3$};
  
  \path[every node/.style={anchor=south,auto=false,align=center}]
        (A) edge (B)
        (B) edge  (C)
        (B) edge (D)
        (D) edge[bend right=30]  node (backarc1) [anchor=south] {} (B)
        ($(D)+(-0.2cm,-0.7cm)$) edge[bend left=30]   node (backarc2) [anchor=north] {} ($(B)+(0.2cm,-0.7cm)$);  

\draw [red] ($(B)!0.5!(D) + (0cm,0.2cm)$) ellipse (2.8cm and 0.9cm);
\draw [red] ($(B)!0.5!(D)$) ellipse (3cm and 1.5cm);
\end{tikzpicture}}%
\end{minipage}
\par
\flushleft
\makebox[0.25\textwidth][c]{(a)}%
\makebox[0.35\textwidth][c]{(b)}%
\makebox[0.35\textwidth][c]{(c)}

\caption{Illustration of the decomposition of a flowchart (coming from a C program) into nested loops. Note that this is not a program in our language but
in a language of guarded assignments, which is, hopefully, self-explanatory. We only present it to illustrate the considerations in 
Section~\ref{sec:FCmotivations}.}
\label{fig:Cprog}
\end{figure}


\section{Loop-Annotated Flowchart Programs}
\label{sec:FCdefs}


\subsection{Program form and informal semantics}

\paragraph{Data}
Our programs operate on a finite (per program) set of variables, each holding  a  number.
The variables are typically denoted by $\X_1,\dots,\X_n$, and a state of the program's storage is thus a vector
$(x_1,\dots,x_n)$. The initial contents of the variables are regarded as input to the program, and their final value as output.
We assume that the only type of data is nonnegative integers.
More generality is possible, e.g., considering signed integers.
This will be left out of the present paper.

\paragraph{Instructions}
\label{sec:atomic-core}

Atomic commands, or \emph{instructions}, modify the values of variables. A \emph{core instruction set} for our work,
corresponding to the instructions of $\Lbjk$, consists of
the instructions

\verb+X := Y+, \quad \verb/X := Y + Z/, \quad \verb+X := Y * Z+, \quad \verb+skip+, \quad \verb+X := **+

\noindent
where \texttt{X, Y} and \texttt{Z} are variable names. The \verb+skip+ instruction is a no-op and could be
replaced with \verb+X := X+.  The last form means ``set \pgt{X} to an unknown value," which of course will have no
upper bound in terms of the input;  it is included because it is useful in abstracting realistic programs.
We remark that, with an eye to abstraction of real programs, we may also use the \emph{weak assignment} forms

$\X \wpass \Y$, \quad $\X \wpass \verb/Y + Z/$, \quad $\X \wpass \verb+Y * Z+$, 

\noindent
which set $\X$ to a non-determined value between 0 and the right-hand side. 
A detailed semantics for the instructions may be found in~\cite{BP:flowcharts-TR}.


\bdfn
A \emph{loop-annotated flowchart-program} (abbreviated to ``a program," when context permits)
 $\prog$ consists of:
\begin{itemize}
\item A finite set of variables 
$\X_1,\dots,\X_n$.

\item A \emph{control-flow graph} (CFG) which is a directed graph
$G_{\prog}=(\locations{\prog},\arcs{\prog})$.
The nodes $\locations{\prog}$ are called  \emph{locations}. 

\item A map $\instr{}$ from CFG arcs to instructions
(we sometimes abuse the term and refer to arcs as ``instructions;" the meaning should
be clear from context).

\item A non-empty set of entry nodes (nodes with no predecessors), $P_{entry}$,
and a non-empty set of terminal nodes (nodes with no successors), $P_{term}$ (while program units in most languages have
a single entry point, the generality of allowing a set of entry nodes is useful in our algorithm).

\item A \emph{loop tree}, described next.
\end{itemize}
\edfn


\bdfn
\label{def:loop-tree}
A loop tree for program $\prog$ is a set $\loops{\prog}$ of 
subsets of $\arcs{\prog}$, called loops, which form
a rooted tree under a relation which we call ``nesting". Loops nested in $L\in \loops{\prog}$ are required to be disjoint, strict subsets of $L$.
The root is the whole CFG.
With each non-root
$L\in\loops{\prog}$ is associated a variable, called its \emph{bound}; technically, we denote by
$\bound{L}$ this variable's index, so that the bound is $\X_{\bound{L}}$. In addition, a ``cut set" $\cutset{L}$ is
provided, which consists of arcs that belong to $L$ but not to its descendants.
 In a valid program, if $a\in L$, instruction $\instr{a}$ must not modify
$\X_{\bound{L}}$. In addition, every cycle $C$ in the CFG must include an arc from the
cutset of the lowest loop $L$ containing $C$.
\edfn

For intuition, we might think of cut-set instructions as maintaining a counter
which ensures that flow passes through such arcs at most $\X_{\bound{L}}$ times.
We make the assumption---without loss of generality---that cut-set arcs represent transitions that do not alter
the values of variables; this condition can always be achieved by adding arcs to the graph. We shall use the symbol $\ensuresymbol$ to
mark these arcs in a diagram.
Note that the notion of ``a loop'' is very flexible. It is a set of arcs, in particular is not
required to be strongly connected (though this would be the natural situation).
One benefit of this flexibility is that this loop information can persist through program transformations.

\paragraph{Semantics} of programs is mostly straight-forward.
The  loop information is interpreted as follows: once a
loop $L$ is entered, at most $B$ cutset arcs of $L$ may be traversed before the loop is exited,
 where $B$ is
the value of the loop bound.  For full details of the definition, see Appendix~\ref{sec:semantics}.

%

\subsection{Growth-rate analysis}
The \emph{polynomial-bound analysis problem} is to find,
for a given program, which output variables
are bounded by a polynomial in the initial values of all variables.
This is the problem we focus on; the following variants can be easily reduced to it:
(1) \emph{feasibility}---find whether all the values generated throughout any computation (rather than outputs only)
are polynomially bounded in the initial values. 
(2)
\emph{Polynomial running time}---find whether
the worst-case time complexity of the program (i.e., number of steps) is polynomial.

\subsection{Flowcharts versus structured programs}

Flowcharts are more expressive than structured programs%
\footnote{This means programs sharing the structure of $\Lbjk$: they include loop commands, commands that branch into two sub-commands,
and atomic commands, and have the single-entry-single-exit property.}
 over the same instruction set,
since they can have complex ``non-structured" control flow;
we now propose a formal argument to support this claim.
It is natural to say that a flowchart $F$ is \emph{strongly equivalent} to a program $P$ if they have the same set of traces, a trace being the sequence of 
atomic commands performed in a computation.
More precisely, let ${\mathcal T}_F(\vec x)$ (respectively ${\mathcal T}_P(\vec x)$) be the set of traces of the flowchart (resp.~structured
program) when the initial state is $\vec x$; say that $F$ and $P$ are \emph{equivalent} if 
${\mathcal T}_F(\vec x) = {\mathcal T}_P(\vec x)$ for all $\vec x$. 
This clearly implies 
$\bigcup_{\vec x} {\mathcal T}_F(\vec x) = \bigcup_{\vec x} {\mathcal T}_P(\vec x)$, an equality that we call
\emph{weak equivalence}.
This last set
is a regular language over the alphabet of instructions. Note that by ignoring the semantics of instructions and considering
them as abstract symbols,
a flowchart is reduced to a finite
automaton (NFA). A structured program can be directly represented by a regular expression over the same alphabet.
By~\cite{Eggan1963}, a flowchart that includes, in a single loop, the 2-node 
digraph
\qquad\rule{0pt}{1.2em}\(
\xymatrix{
\bullet \ar@(dl,ul)[]\ar@/^4pt/[r]& 
\bullet  \ar@(dr,ur)[]\ar@/^4pt/[l]
}
\)\qquad
(labeled with distinct instructions) has the property that any 
weakly-equivalent structured program
has star-height of $2$ at least (i.e., it includes the pattern $(...(...)^*...)^*\,$). Now, the
flowchart program, as it has a single loop, has linear running time, whereas the structured program will have a worst-case running 
time quadratic at least. Hence, they cannot be strongly equivalent. In other words, such an annotated flowchart
has no equivalent structured program
 (though both have polynomial running time; indeed, a
FC-to-\Lbjk transformation that preserves polynomiality follows from our results,
 but it has another drawback---an exponential blow-up in size).



\section{Loop-Annotated Regular Expression Programs}
\label{sec:LAREdefs}


In this section we introduce \emph{Loop-Annotated Regular Expression Programs}, \lare,
which is a program form designed to exploit the analogy of structured programs to regular expressions and flowcharts
to automata.
This language is based on regular expressions but it represents programs in a superset of \Lbjk,
and is endowed with computational semantics. 
However, occasionally we do refer to the standard semantics of a regular expression as describing a set of strings.
To emphasize the former interpretation, we may use the term ``program'' instead of ``expression".

\bdfn
The class of Loop-Annotated Regular-Expression Programs \lare is constructed as follows. First, \emph{atomic}
expressions are:
\be
\item
A set $\instset$ of \emph{basic instructions}, which serve as \emph{symbols} of the expressions. 
\item
An additional, special symbol, $\ensuresymbol$, called the cut-arc symbol.
\item
The empty-string constant $\epsilon$.
\ee
\noindent
Secondly, expressions (atomic or not) can be combined in the following ways
\be
\item
Concatenation: $EF$ where $E$, $F$ are \lare expressions.
\item
Alternation (or non-deterministic choice), $E|F$ where $E$, $F$ are expressions.
\item
Iteration: $E^*$ where $E$ is an expression.
\item
Loop annotation: if $E$ is an expression and $\X_\ell$ a variable, we can form the
expression $[_\ell E]$, provided that $E$ does not include any assignment to $\X_\ell$.
The pair of syntactic elements $[_\ell$ and $]$ is called \emph{loop brackets}.
\ee 

If we ignore the loop brackets, an \lare expression is just a regular expression which generates a set of strings over $\instset$.
We use $\mbox{\lsem{$E$}\rsem}_{str}$ to denote this set.
The loop brackets have a role in defining the computational semantics of \lare.  First, we state a validity requirement: In a \emph{well formed} expression,
every iteration construct $E^*$ must appear inside some pair of loop brackets. Moreover, every string that
$E$ generates must include a $\ensuresymbol$. E.g., $(a(\ensuresymbol b)^*d)^*$ is invalid,
because the expression $(a(\ensuresymbol b)^*d)$ generates the $\ensuresymbol$-free string $ad$.
\edfn
Parentheses will be used to indicate expression structure,
 as usual (also using standard precedence and associativity for operators).

\emph{Semantically}, an \lare program executes a (non-deterministically chosen) sequence of instructions that, viewed as a string over
$\instset$, belongs to the language  $\mbox{\lsem{$E$}\rsem}_{str}$.
However, the sequence also has to satisfy the loop bounds, in the sense
that the number of \ensuresymbol's encountered in a subsequence generated by expression $[_\ell E]$, discounting those
in the scope of any inner pair of brackets, is bounded by $\X_\ell$. For full details of the definition, see Appendix~\ref{sec:semantics}.

The loop brackets make possible the correspondence between
$\lare$ and flowcharts. We take the following result to be intuitive, and we state it without details:

\bthm
An $\lare$ program can be represented as a semantically-equivalent loop-annotated flowchart over the same set of instructions.
\ethm

The semantic equivalence between the two program forms is one of trace semantics---as long as program locations (nodes in the flowchart graph)
are disregarded in the traces (this is also formalized in Appendix~\ref{sec:semantics}).

It should also be pretty obvious that the \emph{structured core language} $\Lbjk$ can be embedded in \lare{}---this is
really just a change of syntax, where, importantly, a loop $\verb/loop X/_\ell \verb/ {C}/$ becomes $[_\ell (\ensuresymbol E)^*]$, where $E$
represents \C.  Hence, in this translation, the iteration operator (star), cut-arc symbol and loop brackets always work together. Generally, this is
not required in \lare, allowing more flexibility, which is required for their equivalence to flowcharts.

\ifconf\else
The subset \lareoneE consists of programs in which every star operator is used in combination with
the $\ensuresymbol$ thus: $(\ensuresymbol E)^*$, and there are no other {\ensuresymbol}s.
Since every iterated segment is required to have a $\ensuresymbol$, changing an \lare program to \lareoneE format may,
at worst, have the effect that  a number greater than one of {$\ensuresymbol$}s are replaced by a single one
(e.g., $(A\ensuresymbol B\ensuresymbol)^* \to (\ensuresymbol AB)^*$), which has no effect on our decision problem
of polynomial bounds (it also affect a polynomial bound by at most a constant factor).
We use this subset sometimes to simplify presentation.
\fi

\begin{figure}[t]
\begin{minipage}{0.4\textwidth}
\begin{Verbatim}[commandchars=\\\{\},codes={\catcode`$=3\catcode`_=8}]
loop $\X_4$ \{
   $\X_3$:=$\X_1+\X_2$;
   $\X_2$:=$\X_3$
\}
\end{Verbatim}
\smallskip
 $[_4 (\ensuresymbol ( \asum{3}{1}{2} \acopy{2}{3} ) \ )^* ]$
\end{minipage}
\hspace{0.1\textwidth}
\begin{minipage}{0.4\textwidth}
\begin{Verbatim}[commandchars=\\\{\},codes={\catcode`$=3\catcode`_=8}]
loop $\X_4$ \{
   choose $\X_2$:=$\X_1$+$\X_4$
   or  $\X_2$:=$\X_2$+$\X_4$;
\}
\end{Verbatim}
\smallskip
 $[_4 (\ensuresymbol ( \asum{2}{1}{4} \:|\: \asum{2}{2}{4} ) \ )^* ]$
\end{minipage}


\caption{
programs in $\Lbjk$ and their expression as \lare, to illustrate the latter.
}
 \label{fig:lareExamples}
\end{figure}


\section{Translating Flowchart Programs into LARE}
\label{sec:FCtoLARE}

We now arrive at the translation of a flowchart program into \lare, which was the point of introducing it. This translation is not
difficult, but since our analysis rests on it, we give it in some detail.
Our algorithm is based on the classical \emph{Rip} algorithm to  transform an NFA into a regular
expression (\cite[Sec.~1.3]{Sipser:TOC:3rd}), which we extend in order to handle loops correctly.
We present the algorithm in some stages, bottom-up, starting with a part which is just as in the NFA algorithm.

\emph{Notation}: the algorithm manipulates a graph which has \lare expressions on each arc (this generalizes
an ordinary flowchart, where each arc carries a single instruction, in the same way the \emph{GNFA} generalizes an NFA
in \cite{Sipser:TOC:3rd}; due to lack of space we do not elaborate).
Denote by $\texttt{Rex}_e$ the expression on the arc $e$ (we may write $uv$ for $e$ when there is a single arc from $u$ to $v$).

\subsection{Ripping a node}
By ``ripping'' a node we remove it from the graph, obtaining a smaller graph with equivalent semantics%

\begin{alg}
$\textsc{RipOne}(g,v)$: \\
accepts a control-flow graph $g$ annotated with \lare expressions, and an internal node $v$ (i.e., $v$ has
both in-going and out-going arcs).
\end{alg}
\begin{enumerate}
\item 
Merge parallel arcs in $g$, if any, by combining the expressions with the alternation operator.

\item 
For every path of length two, $uvw$,
add an arc $e'$ from $u$ to $w$ and let  $\texttt{Rex}_{e'}$ be $\texttt{Rex}_{uv}(\texttt{Rex}_{vv})^*\texttt{Rex}_{vw}$
(if $v$ has no self-loop, we get $\texttt{Rex}_{uv}\texttt{Rex}_{vw}$).

\item Remove $v$ from the graph.
%
\end{enumerate}

\subsection{Contracting a loop}

Next we define a procedure to contract a subgraph $g$, representing a leaf loop (that has no child loops),
by ripping its internal nodes.
We assume that the graph has some entry nodes 
and some exit nodes 
and that
these are the only nodes that connect to the rest of the program.
The effect of the contraction procedure below is that only entry and exit nodes remain; and arcs connect entries to exits.

\begin{alg}
$\textsc{ContractSimple}(g,l)$, where $g$ is a control-flow graph and $l$, an index of a loop variable:
\end{alg}
\begin{enumerate}
\item For every node $v$ which is not an entry or exit node in $g$, perform
\textsc{RipOne$(g,v)$}.
\item For every arc $e$ in this bipartite graph, replace its label
$\texttt{Rex}_e$ by $[_l\,\texttt{Rex}_e\,]$.
\end{enumerate}

\medskip
\noindent
The full \textsc{Contract} procedure handles a loop in the context of a larger flowchart. 
Since loops may in general, share nodes, we first isolate
the loop from the rest of the graph and ensure that it has well-defined entry and exit nodes, by creating new nodes
for this purpose, which we call
virtual entry/exit nodes. More precisely, 
a \emph{boundary node} of a subgraph is a node which is incident to arcs outside the subgraph as well as within.
If $v$ is such a node, it is replaced (in the most general case) by four nodes: $v_{\overline L}$ for paths that stay outside the loop, $v_{in}$ to enter the loop,
$v_{out}$ to exit it, and $v_L$ for paths that go through $v$ but stay inside 
the loop (this may be best understood with a drawing---see Figure~\ref{fig:boundaryNode}).
Note that we are not handling nested loops yet.
The procedure is illustrated by Figure~\ref{fig:contractLoop}.


\begin{alg}
$\textsc{Contract}(L)$
\end{alg}
\begin{enumerate}

\item For every boundary node $v$ of $L$, create 
new nodes $v_L$, $v_{\overline L}$, $v_{in}$ and $v_{out}$.
Any arc incident to $v$ is replaced by two arcs as shown in  Figure~\ref{fig:boundaryNode}).
 
 \item Let $g'$ consist of the subgraph spanned by internal (non-boundary) nodes of $L$, plus, for a boundary node $v$,
 the nodes $v_{in}, v_{out}, v_L$.
 
\item Perform \textsc{ContractSimple}$(g',\X_{\bound{L}})$. This removes all the internal nodes of $L$, leaving only the entry
and exit nodes.
\end{enumerate}

\subsection{Converting a whole program}

The algorithm to convert a whole flowchart program to \lare now follows easily.
The input to the algorithm is, in general, a loop in the loop tree of the flowchart program (Definition~\ref{def:loop-tree}).
To convert the entire program, we apply \textsc{ConvertFC} to the root of the tree.

\begin{alg}
$\textsc{ConvertFC}(L)$
\end{alg}
\begin{enumerate}

\item Perform recursive calls to \textsc{ConvertFC} for all the children of $L$. In these recursive calls,
any virtual nodes created will be shared (i.e., if $v$ is a specific node of $L$ and it has incident arcs in two
subloops, there will still be only one node $v_{in}$ and only one $v_{out}$. This subtlety arises because our loops are defined to be edge-%
disjoint but not vertex-disjoint).

\item Perform \textsc{ContractSimple}$(L,\X_{\bound{L}})$. 
\end{enumerate}

As an exception, for the root loop we simplify \textsc{Contract} by not creating virtual nodes. This works because it has
entry nodes 
and exit nodes 
by definition, and we do not want to alter their identity.

\begin{figure}[t]
\begin{tikzpicture}[-triangle 45,auto,node distance=1.7cm]
\tikzset{every state/.style={minimum size=1.8em,inner sep=0pt}}
	\node [state] (n1)  {$A$};
	\node [state] (n3) [below right of=n1] {$v$};
	\node [state] (n2) [below left of=n3] {$B$};
	\node [state] (n4) [above right of=n3] {$C$};
	\node [state] (n5) [below right of=n3] {$D$};
    \path[decoration={snake,amplitude=.4mm,segment length=2mm}]
		(n1) edge  node [anchor=south]  {$a$} (n3)
		(n3) edge  node [anchor=east]  {$b$} (n2)
		(n3) edge[decorate]  node [anchor=south]  {$c$} (n4)
		(n5) edge[decorate]  node [anchor=east]  {$d$} (n3);
\end{tikzpicture}
\raisebox{1cm}{$\Longrightarrow$}
\begin{tikzpicture}[-triangle 45,auto,node distance=1.7cm]
\tikzset{every state/.style={minimum size=1.8em,inner sep=0pt}}
	\node [state] (n1)  {$A$};
	\node [state] (n3) [below right of=n1] {$v_{\overline L}$};
	\node [state] (n2) [below left of=n3] {$B$};
	\node [state] (v1) [above right of=n3] {$v_{in}$};
	\node [state] (v3) [below right of=v1] {$v_{L}$};
	\node [state] (n4) [above right of=v3] {$C$};
	\node [state] (n5) [below right of=v3] {$D$};
	\node [state] (v2) [below right of =n3] {$v_{out}$};
	\path
		(n1) edge  node [anchor=south]  {$a$} (n3)
		(n3) edge  node [anchor=east]  {$b$} (n2)
		(n1) edge  node [anchor=south]  {$a$} (v1)
		(v2) edge  node [anchor=east]  {$b$} (n2);
	\path[decoration={snake,amplitude=.4mm,segment length=2mm}]
		(v1) edge[decorate] node [anchor=south]  {$c$} (n4) 
		(n5) edge[decorate] node [anchor=south]  {$d$} (v2)
		(v3) edge[decorate]  node [anchor=south]  {$c$} (n4)
		(n5) edge [decorate] node [anchor=south]  {$d$} (v3);
\end{tikzpicture}
\caption{Transforming a boundary node $v$. The curly arcs belong to the loop under consideration.
Arc labels $a,b,\dots$ represent instructions.}
\label{fig:boundaryNode}
\end{figure}

\begin{figure}[tb]
\noindent
\begin{tikzpicture}[-triangle 45, auto, node distance=1.8cm, thick,baseline=(current bounding box.north)]
\tikzset{every state/.style={minimum size=1.8em,inner sep=0pt}}
  \node[state] (A)   {$A$}; 
  \node[state] (B) [right of=A]  {$B$};
  \node[state] (C) [right of=B]  {$C$};
  \node[state] (F) [right of=C]  {$F$};

  \path[every node/.style={anchor=south,auto=false,align=center}]
        (A) edge[bend left] node [anchor=south] {$a$}                 (B)
        (B) edge[bend left] node [anchor=south] {$b$}                 (A)
        (B) edge [bend right,decorate,decoration={snake,amplitude=.4mm,segment length=2mm,post length=1mm}] node [anchor=north] 
{$\ensuresymbol$} 
   (C)
        (C) edge [bend right,decorate,decoration={snake,amplitude=.4mm,segment length=2mm,post length=1mm}] node [anchor=south]{$d$} (B)
        (C) edge[bend right] node [anchor=south] {$e$}                 (F)
        (F) edge[bend right] node [anchor=south] {$f$}                 (C)
        ;
\end{tikzpicture}

\noindent
\iflncs
\def\nodedistforthispic{3cm}
\def\twnodedistforthispic{6cm}
\else
\def\nodedistforthispic{4cm}
\def\twnodedistforthispic{8cm}
\fi
\begin{tikzpicture}[-triangle 45, auto, node distance=\nodedistforthispic, thick,baseline=(current bounding box.north)]
\tikzset{every state/.style={minimum size=1.8em,inner sep=0pt}}
  \node[state] (A)   {$A$}; 
  \node[state] (B) [right=1.5cm of A]  {$B_{\overline L}$};
  \node[state] at ($(B)+(2.5,1)$) (Bin)   {$B_{in}$};
  \node[state] at ($(B)+(2.5,-1)$) (Bout)  {$B_{out}$};
  \node[state] (Cin) [right of=Bin]  {$C_{in}$};
  \node[state] (Cout) [right of=Bout]  {$C_{out}$};
  \node[state] (C) [right=\twnodedistforthispic of B]  {$C_{\overline L}$};
  \node[state] (F) [right=1.5cm of C]  {$F$};

  \path[every node/.style={anchor=south,auto=false,align=center}]
        (A) edge[bend left] node [anchor=south] {$a$}                 (B)
        (B) edge[bend left] node [anchor=south] {$b$}                 (A)
        (A) edge [bend left] node [anchor=south] {$a$}           (Bin)  
        (Bout) edge [bend left] node [anchor=north] {$b$}        (A)  
        (C) edge[bend right] node [anchor=south] {$e$}                 (F)
        (F) edge[bend right] node [anchor=south] {$f$}                 (C)
        (Bin) edge node [anchor=east] {$\lbrack_1(\ensuresymbol d)^*\rbrack$} (Bout)
         (Bin) edge [bend right] node [near end,anchor=north east,xshift=0.3cm]
           {$\lbrack_1 \ensuresymbol\left(d\ensuresymbol\right)^*\rbrack$}             (Cout)
        (Cin) edge [bend right] node [near start,anchor=south east,xshift=0.3cm] 
           {$\lbrack_1 d\left(\ensuresymbol d\right)^*\rbrack$}   (Bout)
        (Cin) edge node [anchor=east]
           {$\lbrack_1\left(d \ensuresymbol\right)^*\rbrack$}              (Cout)
        (Cout) edge [bend right] node [anchor=north] {$e$}         (F)  
        (F) edge [bend right] node [anchor=south] {$f$}            (Cin)  
        ;
\end{tikzpicture}

\caption{A flowchart with a loop $\{B,C\}$ (above) and the result of contracting this loop.
We assume that the bounding variable for the loop is $\X_1$. Note that nodes $B_L$, $C_L$ are not present since contraction of the loop has removed them.
}
\label{fig:contractLoop}
\end{figure}


\section{The Growth-Rate Analysis Algorithms}
\label{sec:LAREalgo}

In this section we reach the main theoretical contribution: a polynomial-time decision procedure for
polynomial growth rates. We implement the idea of extending results from well-structured programs
to arbitrary control-flow graphs by exploiting the translation of graphs to regular expressions. 
First, we extend the analysis of~\cite{BJK08} to {\lare}
(the extension is easy, but for the sake of completeness we give the algorithm in a self-contained form.
This version is based on~\cite{B2010:DICE} rather than~\cite{BJK08}).

For the rest of this article, let $\inds = \{1,\dots,n\}$ be the set of variable indices.

\subsection{Analysis of $\lare$ commands}

We define an interpretation of \lare commands over the domain of \emph{dependency sets}.
By applying this interpretation to a command, we find how the magnitude of
the values of each at the end of the computation depends on the initial values.
 
\subsection{Dependencies and dependency sets}

\bdfn
The set of \emph{dependency types} is
$\deptypes =  \{1,1^+,2,3\}$, with order $1 < 1^+ < 2 < 3$, and maximum operator
$\sqcup$.  We write $x\simeq 1$ for $x\in\{1,1^+\}$.
\edfn

\noindent
The following verbal descriptions  may give intuition to the meaning of dependency types:

$1=$\emph{identity dependency}, \par
$1^+=$\emph{additive dependency}, \par
$2=$\emph{multiplicative dependency}, \par
$3=$\emph{exponential dependency} (more precisely, super-polynomial). 

\bdfn
The set of \emph{dependencies} is $\depsdice$,
which is the union of two sets: \\ (1) The set of \emph{unary dependencies},
isomorphic to $\inds\times\deptypes\times\inds$. The notation for an element is
\ $\unarydep{i}{\delta}{j}$. \\ (2) The set of \emph{binary dependencies},
isomorphic to 
$\inds\times\inds\times\inds\times\inds$, notated
\ $\binarydep{i}{j}{k}{\ell}$. 
\edfn

Intuitively, binary dependencies represent conjunctions of unary dependencies. The fact that it is
necessary and sufficient to handle conjunctions of pairs of unary dependencies (but not of larger sets) 
is a non-trivial property which is key to this algorithm's correctness.

\bdfn A \emph{dependency set} is a subset of $\depsdice$, with the proviso that
a binary dependency $\binarydep{i}{j}{k}{\ell}$ may appear in the set only if
it also includes $\unarydep{i}{\alpha}{k}$ and $\unarydep{j}{\beta}{\ell}$,
for some $\alpha, \beta \simeq 1$, and $i\ne j \lor k\ne \ell$.
\edfn

The function $\addbdeps$ adds to a dependency set all binary dependencies that can be added according to
the above rule. That is:

\noindent
For a dependency set $S$, we let
\(
 \addbdeps(S) \eqdef S \cup \{ \binarydep{i}{j}{k}{\ell} \mid\ \unarydep{i}{\alpha}{k}\in S \,\land\, \unarydep{j}{\beta}{\ell}
                   \in S,\ \text{$\alpha, \beta \simeq 1$} \}.
\)
We then define the \emph{identity} dependency set is 
\(
 \iddep \eqdef \addbdeps (\; \{ \unarydep{i}{1}{i} \mid i\in\inds \} \;).
\)

\subsection{Interpretation of \lare}
\label{sec:lareAnalysis}

To give a dependency-set semantics to an \lare program $\lareprog$, 
which we denote by $\depsempar{\lareprog}$,
we give a semantics to every symbol and every operation. 

\subsubsection{Symbols}
 The symbols of \lare are atomic instructions, which update the state, and are supposed to be associated
with some dependency sets. In order to justify our claim to a complete solution for the core instruction set,
we give the dependency sets corresponding to these assignment instructions:

{
\renewcommand{\fbox}[1]{\mbox{#1}} 
 \begin{align*}
\depsempar{{\askip}} &= \iddep\\
\depsempar{{\acopy{r}{s}}} &= \addbdeps (\; \{ \unarydep{s}{1}{r} \} \cup \{ \unarydep{i}{1}{i} \mid i\ne r \} \;)\\
\depsempar{{\asum{r}{s}{t}}} &= \addbdeps (\; \{ \unarydep{s}{1^+}{r}, \unarydep{t}{1^+}{r} \} 
          \cup \{ \unarydep{i}{1}{i} \mid i\ne r \} \;) & \text{when $s\ne t$}\\
\depsempar{{\asum{r}{s}{s}}} &= \addbdeps (\; \{ \unarydep{s}{2}{r} \} 
          \cup \{ \unarydep{i}{1}{i} \mid i\ne r \} ) \\
\depsempar{{\amul{r}{s}{t}}} &= \addbdeps (\; \{ \unarydep{s}{2}{r}, \unarydep{t}{2}{r} \} 
          \cup \{ \unarydep{i}{1}{i} \mid i\ne r \} \;) 
 \end{align*}
}

The atomic command \verb+X := **+ (set to unbounded value) is not directly representable. We handle it as 
\verb+X := HUGE+ where \pgt{HUGE} is a special variable. A weak assignment $\X \wpass \texttt{e}$ is abstracted the same 
as the corresponding deterministic assignment \verb+X := e+. 

\subsubsection{LARE operators}

\paragraph{Alternation} is interpreted by set union:
\[
 \depsempar{E|F} = \depsempar{E} \cup \depsempar{F}
\]

\paragraph{Concatenation} is interpreted as a component-wise product of sets:
\[
 \depsempar{EF} = \depsempar{E} \cdot \depsempar{F}
\]
where the product of dependencies is given by

\bdfn[dependency composition] \label{def:prod}
\renewcommand{\arraystretch}{2} 
\[
\begin{array}{ccc}
   (\unarydep{i}{\alpha}{j}) \cdot (\unarydep{j}{\beta}{k}) &=& 
   \multicolumn{1}{l}{\longunarydep{i}{\alpha\sqcup\beta}{k}} \\  
   (\unarydep{i}{\alpha}{j}) \cdot (\binarydep{j}{j}{k}{k'}) &=& (\binarydep{i}{i}{k}{k'}),
\quad\text{provided $\alpha \simeq 1$} \\
    (\binarydep{i}{i'}{j}{j}) \cdot (\unarydep{j}{\alpha}{k}) &=& (\binarydep{i}{i'}{k}{k}),
\quad\text{provided $\alpha \simeq 1$} 
\\
   (\binarydep{i}{i'}{j}{j'}) \cdot (\binarydep{j}{j'}{k}{k'}) &=& 
\left\{\begin{array}{cl}
\binarydep{i}{i'}{k}{k'},  &  \text{ if $i\ne i'$ or $k\ne k'$} \\
\unarydep{i}{2}{k},  &  \text{if $i=i'$ and $k= k'$}
\end{array}\right.
 \end{array}
\]
\edfn

\noindent
The last sub-case in the definition 
handles commands whose effect is to double a variable's
value by making two copies of it and adding them together.

\paragraph{Iteration} To define the interpretation of the star operator,
we first define
$LFP(S)$
to be the least fixpoint (under set containment) of the function
\(
f(X)= \iddep \cup X \cup (X\cdot S) \,.
\)

%
Finally, we define the so-called \emph{loop correction} operator, 
which represents the \emph{implicit dependence} of variables updated in a loop on the loop bound.
First, we define it for single dependencies:
\begin{equation}
\label{def:LC}
\begin{alignedat}{2}
& LC_\ell (\unarydep{i}{1^+}{i}) && =
   \{ \unarydep{\ell}{2}{i} \}   \\
& LC_\ell (\unarydep{i}{2}{i}) && =
  \{ \unarydep{\ell}{3}{i}  \} \\
& LC_\ell (\Delta) && = \{\,\} \qquad \text{for all other $\Delta\in\depsdice$,}
\end{alignedat}
\end{equation}
and then extend it to sets by $LC_\ell(S) = S\cup \bigcup_{D\in S} LC_\ell(D)$ (note that we define it so that
$LC_\ell(S)\supseteq S$).
Using these definitions,  let $F = LFP(\depsempar{E})$, then
\[
 \depsempar{E^*} = LC_\ell(F)  \cdot F
\]
where $\ell$ is the index of the bounding variable for the closest enclosing
bracket construct.

In this analysis, the brackets themselves do not imply any computation in the abstract semantics.
Their role is just to determine $\ell$ in the above rule for loop correction.

\bex
Consider the loop in Figure~\ref{fig:lareExamples} (left).
It can be expressed as an \lare command of the form $\lbrack_4 (\ensuresymbol E)^* \rbrack$,  where $E$ represents the loop body.
The set of \emph{unary} dependencies for $E$ is
\[
 \unarydep{1}{1}{1}, \quad
 \unarydep{1}{1^+}{3},\quad
 \unarydep{2}{1^+}{3}, \quad
 \unarydep{1}{1^+}{2}, \quad
 \unarydep{2}{1^+}{2}, \quad
 \unarydep{4}{1}{4}
\]
Binary dependencies include all pairs of the above. Now consider the product
$\depsempar{E} \cdot \depsempar{E}$; by the last case in Definition~\ref{def:prod}, it includes 
$\unarydep{1}{2}{2}$. This demonstrates the role of this definition in expressing the fact that when the loop
is iterated, a multiple of the initial value of $\X_1$ is accumulated in $\X_2$. Now, consider the dependency
$\unarydep{2}{1^+}{2}$, which may be interpreted as stating that $\X_2$ is an accumulator.
Since this dependency exists in the closure $F= LFP(\depsempar{E})$, this will produce
an additional dependency when we apply the $LC$ operator, specifically, it generates $\unarydep{4}{2}{2}$.
Indeed, $\X_2$ accumulates a multiple of $\X_4$ (in fact, $\X_4\cdot \X_1$; but we do not record the precise
expression). Finally, when computing $LC_4(F)\cdot F$, we compose
$\unarydep{4}{2}{2}$ with $\unarydep{2}{1^+}{3}$ obtaining $\unarydep{4}{2}{3}$, which reflects the
flow of the product $\X_4\cdot \X_1$ to $\X_3$; similarly, we obtain $\unarydep{1}{2}{3}$.
\eex

\bex
To illustrate the role of binary dependencies, let us compare the loop just analyzed with 
Figure~\ref{fig:lareExamples} (right).
Here, in the loop's body, the unary dependencies $\unarydep{1}{1^+}{2}$ and 
$\unarydep{2}{1^+}{2}$ exist, but not their conjunction (because they happen in alternative execution paths),
and the result $\unarydep{1}{2}{2}$ does not arise.
\eex

\subsection{Correctness and complexity}
\label{sec:lare-analysis-properties}

Our correctness claims state that the results of this analysis provide, for every variable,
either a polynomial upper bound or an exponential lower bound.
This assumes the core instruction set; if additional instructions are added, one of these two soundness claim
may be compromised, depending on how closely the instruction is modeled by dependency sets
(note that, interestingly, our result means that the core instruction set cannot express a computation
with super-polynomial-sub-exponential growth rate).

\bthm \label{thm:poly-correct}
Let $E$ be a well-formed \lare command, using the core instruction set.
If $\unarydep{i}{3}{j}\in \depsempar{E}$ for some $i$, then the values 
that $\X_j$ can take at the end of an execution of $E$ grow, in the worst case, at least exponentially
in terms of the initial variable values. If there is no such dependency,  a polynomial bound on the final value of $\X_j$ exists.
\ethm

The proof of this theorem follows by showing that the dependency sets computed by our algorithm are all sound with
respect to two interpretations: a lower-bound interpretation and an upper-bound interpretation. 
The interpretations are defined in terms of a \emph{trace semantics} of the program.
Due to the complexity of the proofs, they are deferred to appendices.

As to \emph{complexity}, we 
claim that a straight-forward implementation of the algorithm is polynomial-time in the size of the program. The reason is
that the size of a dependency set is polynomially bounded in the number of variables; most operations on such sets
(union, composition) are clearly polynomial-time, the only non-trivial issue being the
fixed-point computation for analyzing $E^*$. However, this is polynomial time because the height of the semi-lattice
of dependency sets is polynomial. 

\subsection{Analysis of flowchart programs}

The principle of the algorithm is to convert the source program to 
\laregen program and perform the above analysis on the result. However, doing it
in two steps, as just described, is
inefficient, because converting an NFA to a regular expression has exponential cost
(in fact, it is known that the \emph{size} of the regular expression cannot be polynomially bounded).

To obtain an efficient algorithm we use the principle of function
fusion~\cite{Chin:1992:fusion}, which basically means to eliminate intermediate structures when
composing functions.  
We fuse together the functions of
conversion from \emph{FC} to
\laregen and  analysis of the \laregen program. The fused algorithm does not generate an expression,
but directly computes its abstract semantics. Hence,
it works on a graph with dependency sets as arc labels, 
rather than \lare expressions, and applies operations in the dependency-set domain instead of syntactic operations
on expressions.

The fused algorithm runs in polynomial time. This follows by bounding the costs of 
the steps---applications of abstract operators---%
and the number of such operations that occur (as they dominate the running time).
The complexity of abstract-domain operations is polynomial, see Section~\ref{sec:lare-analysis-properties}.
For the number of operations, we consider the size of the graph manipulated by the algorithm, which is
polynomial in the size of the original flowchart plus the number of additional nodes generated throughout
the algorithm, due to splitting of boundary nodes. Since such splitting only occurs once per boundary node per loop,
we have a polynomial bound in terms of the original graph.
	


\section{Related Work}

Growth-rate analysis is clearly related to the large body of work on static program analysis for discovering resource
consumption (in particular, running time). But since we focus on decidable cases, based on weak languages,
it is also related to work in Computability and Implicit Computational Complexity. Thus, there is a lot of 
work that could be mentioned, and this section will only give a brief overview and a few representative citations.
 However, we will dwell a bit longer
on recent work that appears interestingly related.

Meyer and Ritchie~\cite{MR:67} introduced
the class of \emph{loop programs}, which only has definite, bounded loops, so that \emph{some}
 upper bound on their
complexity can always be computed. Subsequent work~\cite{KasaiAdachi:80,
KN04,NW06,JK08}
attempted to analyze such programs more precisely;
most of them proposed syntactic criteria, or analysis algorithms, that are
sufficient for ensuring that the program lies in a desired class (say, polynomial-time programs),
but are not both necessary and sufficient: thus, they do not address
the decidability question (the exception is~\cite{KN04} which has a decidability result for a ``core" language).  
As already mentioned, \cite{BJK08} introduced weak bounded loops (such that can exit early) into the loop language,
plus other simplifications, and obtained decidability of polynomial growth-rate. Regarding the necessity of these simplifications,
\cite{BK11} showed undecidability for a language that can only do addition and definite loops (that cannot exit early).

Results that characterize programs in a way sufficient (but not necessary) to have a certain complexity resonate with the
area called Implicit Computational Complexity (ICC), where one designs languages or program classes for capturing
a complexity class; this was the goal in~\cite{KasaiAdachi:80}. Later the approach seems to have focused on
functional languages~\cite{BC:92,Goerdt1992,Hofmann:iandc2003} and term rewriting systems~\cite{BCMT:01}. 

Among related works in static program analysis, mostly 
pertinent are works directed at obtaining \emph{symbolic,
possibly asymptotic, complexity bounds} for programs (in a high-level language
or an intermediate language) under generic cost models (either unit cost or a more flexible, parametrized
cost model).
A symbolic bound could be an expression like $2\pgt{x}+10$, which may be more or less accurate, but to answer a simple
binary question like ``is there a polynomial bound" is not considered sufficient in this area
(though one could argue that weeding out the super-polynomial programs should be of interest).
Characteristic to  this area is the fact that decidable subproblems have rarely been studied.

Wegbreit~\cite{Wegbreit:75} presented the first, and very influential, system for automatically analysing
a program's complexity. His system analyzes first-order LISP programs;
broadly speaking, the system transforms the program into a set of recurrence equations for the complexity which
have to be solved.
Subsequent works along similar lines included \cite{ACE,Rosendahl89}
and more recently \cite{Benzinger01
} for functional programs,
\cite{DebrayLin93
} for logic programs, and~\cite{
Albert-et-al:TCS:2011}  
for JBC.  

Other techniques for resource/complexity analysis of realistic programming languages include
abstract interpretation (for functional programs: \cite{SAFE-FOPARA2010}), 
counter instrumentation~\cite{
SPEED-POPL09} and type systems~\cite{Jost:2010,Hermengildo-atal.ICLP14}.


For programs where loops are not explicitly bounded, there is an obvious connection of finding bounds to proving
termination. So techniques of termination proofs have migrated into bound computation. One example is
size-change termination~\cite{leejonesbenamram01,BA:mcInts} used in~\cite{ZGSV-sas11
},
and another is linear and lexicographic ranking functions~\cite{ADFG:2010}.
This latter work, in particular, has inspired our notion of annotated flowcharts,
as described in Section~\ref{sec:FCmotivations}. By examining~\cite{ADFG:2010} one can see that
they implicitly construct the loop tree. However, once this is done, they can compute a global bound only
from loop bounds which are linear in the program's input.
Thus the method precludes programs where a loop bound
is a non-linear function of the input, possibly computed at a previous (or enclosing) loop. Recently, 
Brockschmidt et al.~\cite{Giesl:TACAS2014} presented an analysis algorithm, called \koat, that deals with this challenge by
alternating bound analysis (based on ranking functions) and size analysis (basically, analyzing the growth rate of variables);
this solution seems to be very interestingly related to our work. We note the following points:
\begin{compactitem}
\item
\koat handles unannotated programs in a realistic language, and it searches for ranking functions as part of the 
algorithm. It does not abstract programs into a weak programming language. No decidability result is claimed, nor
does this seem to be a goal.
\item
\koat produces explicit bounds, taking constants into account, while we only considered the decision problem of
polynomial growth rate, and the goal was to answer it precisely. 
\item
Interestingly,~\cite{Giesl:TACAS2014} use a ``data-flow graph" in their algorithm, while we used such a graph in
the proof
(in fact, a similar graph was already used in~\cite{JK08}). 
\end{compactitem}
We can make a theoretically-meaningful comparison by focusing on the intersection of the two problems solved:
namely, we can look only at our restricted programming language (which can be easily compiled into the input language
of \koat). For such programs we recognize all polynomial programs, while \koat does not (but it provides explicit bounds
when it does).   This suggests an idea for further research, namely to make a closer comparison, and possibly
merge the techniques.  

\section{Conclusion and Open Problems}

This works addressed the decidability of a growth-rate property of programs, namely polynomial growth, in a weak
programming language. Extending previous work that addressed compositionally-structured programs
we have presented an analysis that works for flowchart programs
with possibly complex control-flow graphs, provided with hierarchical loop information.
We propose this program form as a way to extend program analysis algorithms from structured programs to less-structured
ones. We prove that the polynomial growth problem is PTIME-decidable for our class of flowchart programs,
with a restricted instruction set.

Unlike typical work in static analysis of programs (going by names such as \emph{resource analysis} or 
\emph{cost analysis}), our algorithm does not output full expressions for the complexity bounds.
In principle, one could extend our algorithm to produce such bounds, since their calculation is implicit
in the proof. We opted not to do it, since we focus on the problem that we can solve completely.
We acknowledge that if we produce explicit polynomials, they will not be tight in general.
It is an interesting problem, theoretically, to research the problem of computing precise bounds.
Another theoretical challenge is to extend the language, e.g., precisely analyzing a larger set of instructions,
or adding recursive procedures.
In practice, one works with full languages and settles for sound-but-incomplete solutions, but we hope that our line
of research can also contribute ideas to the more practical side of program analysis.

\appendix

\section{Formal Semantics of FC and LARE}
\label{sec:semantics}

This appendix gives additional details on the formal semantics  of both FC programs and LARE programs.
This is required for justifying the correspondence of the two formalisms (which we just state: we skip the correctness proof 
of the conversion algorithm, which is quite trite once the definitions are in place), and for
correctness proof of the growth-rate analysis for LARE (described in the following two appendices).

In both cases we state the semantics in terms of \emph{traces}, sometimes called transition sequences.
Those for FC are ``more concrete'' in that they involve the nodes and arcs of the CFG.

\subsection{Semantics of FC}
\label{sec:FCsemantics}

Consider an FC program $\prog$ with variables $\X_1,\dots,\X_n$,
and control-flow graph $G_{\prog}=(\locations{\prog},\arcs{\prog})$.  For
$a\in\arcs{\prog}$,
write $a:P\to Q$ if $P$ is its source location and $Q$ its target.

\bdfn[states] 
The set of states of $\prog$ is $\states{\prog} = \locations{\prog} \times \nats^n$,
such that $s=(P,\vec x)$ indicates that the program is at location $P$ and
$\vec x$ specifies the values of the variables.
\label{FC-state}
\edfn

\bdfn[transitions] \label{def:FC-trans}
A transition is a pair of states, a \emph{source state} $s$ and a \emph{target
state} $s'$,
related by an instruction $a$ of $\prog$.   More precisely:  ${a:P\to Q}$,
$s=(P,\vec x)$,  $s'=(Q,{\vec x}')$,
and the relation of ${\vec x}'$ to $\vec x$ is determined by the semantics of 
$\instr{a}$.
When this holds, we write $\trans{a}{s}{s'}$ . The set of transitions is $\fctransitions{\prog}$.
\edfn


\bdfn[transition sequence]
A {\em transition sequence\/} (or trace) of $\prog$ is a finite sequence of consecutive state
transitions
$\tilde s = \trans{a_1}{s_{0}}{s_1}\dots\trans{a_t}{s_{t-1}}{s_t}$,
where the instruction sequence $a_1 a_2 \dots a_t$ corresponds to a CFG path.
We refer to the arcs of the path as \emph{the arcs of $\tilde s$}.
The set of traces is denoted $\fctseqs{\prog}$.
\label{trans_seq}
\edfn



The definition of a transition sequence does not take the loop bounds
into account.  
Thus it allows for sequences which do not respect the bounds. 
To enforce the bounds, we introduce the next definition.

\bdfn[properly bounded] \label{def:FCproperlybounded}
For a {\em transition sequence\/} $\tilde s$, let $L\in \loops{\prog}$ be the
smallest loop that includes all arcs of $\tilde s$ (the
smallest enclosing loop). Let $L^\circ$ be $L$ minus any nested loop.
Then, $\tilde s$ is \emph{properly bounded} if the following conditions hold:
\begin{compactenum}
\item  If $L$ is not the root,  let $\ell = \bound{L}$; then 
the number of occurrences in $\tilde s$ of any $\ensuresymbol$ arc from $L^\circ$
is at most  the value of $x_\ell$ (which does not change
throughout $\tilde s$). 
\label{def:pb:1}
\item If $L$ is the root, any $a\in L^\circ$ occurs at most once.
\item Every contiguous subsequence of $\tilde s$ is properly bounded.
\label{def:pb:3}
\end{compactenum}
\edfn



We say the properly-bounded
transition sequence is \emph{a run of loop $L$} when $L$ is the smallest
enclosing loop, as in the definition.
We say that a transition sequence is \emph{complete} if it starts with an entry node of the flowchart
and ends with an exit node.

We let $\fctseqsbnd{\prog}$ denote the set of properly-bounded transition sequences for $\prog$. 

\subsection{Semantics of LARE}

In the semantics of LARE programs, states only describe the values of variables, but not a program location (we may call them \emph{pure states} when distinction
is important).
The set of pure states $\states{}$ is related to $\states{\prog}$ by an obvious abstraction relation.
The evolution of such states is described by (pure) transitions:

\bdfn[pure transitions] \label{def:LAREtrans}
A pure transition is a pair of states, a \emph{source state} $s$ and a \emph{target
state} $s'$,
related by an instruction $a$ out of the vocabulary of atomic instructions.
We write this as $\trans{a}{s}{s'}$ . We presume that a transition correctly reflects the semantics of the atomic instruction.
\edfn

\bdfn[traces]
A {\em trace\/} is a finite sequence of consecutive transitions
$\trans{a_0}{s_{0}}{s_1}\dots\trans{a_{t-1}}{s_{t-1}}{s_t}$. 
The function  $\erase{\sigma}$ removes the $\ensuresymbol$ transitions from the trace $\sigma$,
which is valid because this instruction does not change the state. We define $\ecount{\sigma}$ to be the number of 
$\ensuresymbol$'s in $\sigma$.
The set of all traces is denoted by $\tseqs$ (the number of variables is tacitly assumed to be fixed).
We write $\tseq{s}{\sigma}{s'}$ to indicate that $s[0]=s$ and $s[t]=s'$. 

Concatenation of finite traces $\lambda, \rho$ 
is written as $\lambda \concat \rho$ and requires the final state of $\lambda$ to be the initial state of $\rho$.
As a special case, we denote an \emph{empty} sequence by $\epsilon$ and define $\epsilon\concat\rho = \rho\concat\epsilon
= \rho$ for any $\rho$.
\edfn

We define the \emph{trace} semantics $\tssempar{E}$ of an expression $E$ by structural induction.
First, recall that each \emph{symbol} corresponds to a single instruction. 
 For our core instruction set, semantics should be obvious, so we skip the details.
%
As for \emph{composite programs}, we have
\begin{align*}
 \tssempar{E|F} &= \tssempar{E} \cup \tssempar{F}
\\
 \tssempar{EF} &= \tssempar{E} \concat \tssempar{F}
\end{align*}
where the last operation is, naturally, the component-wise concatenation of $E$ and $F$:
\[
 L \concat R \eqdef \{ \lambda\concat\rho \mid \lambda\in L, \rho\in R, \text{and $\lambda\concat\rho$ is defined}\}.
\]

Finally, for the looping constructs, we define $\tssempar{E^*}$ to be the reflexive-transitive closure of $\tssempar{E}$ under
the concatenation operation $\concat$, and
\[
 \tssempar{ \,[_\ell E ]\, } = \{ \erase{\sigma} \mid \sigma\in \tssempar{E}, \ecount{\sigma}\le (\sigma[0])_\ell \} \,.
\]

\subsection{Correspondence of the semantics}

Let $\prog$ be a flowchart program. Let $\fctseqs{\prog}$ be the set of properly-bounded traces for $\prog$.
As traces of \lare do not involve locations, we define 
the function $\addloc(S, P, Q)$ that inserts locations
in traces $\tilde s \in S$ so that the initial location is $P$, the final location is $Q$, and intermediate locations are the anonymous
location $\bullet$.
We also define a converse function $\droploc(S, P, Q)$, where $S$ is a set of flowchart traces,
 which replaces the program locations
in every $\tilde s \in S$ (except the first location and the last) by $\bullet$, provided the trace starts at $P$
and ends at $Q$ (otherwise, it is ignored).
Then the correspondence of a \lare $E$ to a flowchart program $\prog$ with entry point $P$ and exit $Q$ is expressed by the equation:
\[
\droploc(\fctseqsbnd{\prog},P,Q) = \addloc(\tssempar{E}, P, Q) \,.
\]

\section{Preliminaries for the Proofs}
\label{sec:proof-prelim}

This section includes some preliminaries for the correctness proof of the polynomial-bound analysis (or rather 
proofs: in the next section, we prove that the analysis provides sound lower bounds on the worst-case growth,
 and in the next, that it provides sound upper bounds). Thus we have a sound and complete decision procedure
 for the problem of polynomial growth rate. The proofs in this paper are short presentations, while full details
 can be found in the technical report~\cite{BP:flowcharts-TR}.



Our analysis of $\lare$ programs (Section~\ref{sec:LAREalgo})
 involves an operation (loop correction) that refers to the bounding variable of
the enclosing loop. In order to make the analysis fully compositional (which is easier to reason about),
we avoid the need to ``peek" at the enclosing context by introducing a dummy variable for ``iteration count,"
denoted by $x_{n+1}$ (we may call it also ``the iteration variable", note however that it is
just a place-holder, later to be replaced).  We thus write $\inds_E = \{1,\dots,n+1\}$ for the extended set of variable indices
and extend the notation for
dependencies to $\inds_E$. To the interpretation of all atomic programs we add $\unarydep{n+1}{1}{n+1}$,
and the loop corrector uses $x_{n+1}$ rather than $x_\ell$ (we thus have $LC_{n+1}$ and not $LC_\ell$).
 The bracket construct now has a non-trivial interpretation, namely
let $\subst{\ell}{n+1}{S}$ be the result of substituting any dependency  $\unarydep{n+1}{\delta}{j}$ by $\unarydep{\ell}{\delta}{j}$ in the dependency
set $S$.
Then
\[
 \depsempar{\,[_\ell E ]\,} = \subst{\ell}{n+1}{\depsempar{E}} \,.
\]

We further add to dependencies a property called \emph{color}: black is the default
color and red is special. A dependency is given red if and only if
 it is of the form $\unarydep{n+1}{\delta}{i}$ with $i\ne n+1$.

Some useful observations are:
\begin{inparaenum}[(i)]
\item
In any derivation of dependencies for a LARE program, $LC_{n+1}(D)=\emptyset$ whenever $D$ is red.
\item
In any composition $D_1\cdot D_2$ in a derivation, at most one of $D_1$ and $D_2$ is red.
\item
The type of a red dependence is always 2 or 3.
\item
A black dependence having $n+1$ as source index must be
the identity dependence $\unarydep{n+1}{1}{n+1}$.
\end{inparaenum}


\paragraph{Conventions and notations}
%
For $\vec x = (x_1,\dots,x_n)$,
$\xmin$ is $\min\{ x_i\}$.
The relation $\vec x \spge{} t$
means: $\xmin\ge t$.


\section{Lower-Bound Soundness}
\label{sec-lower-bound}

This section proves soundness of the lower-bound aspect of polynomial-bound analysis,
leading to the conclusion that $\unarydep{x}{3}{y}$ indicates certain exponential growth.
This is the more intuitive interpretation of our abstract domain: we interpret every dependency that
we derive as an indication of something that \emph{can} happen.  In a sense, the heart of the proof is
the proper definition of the concrete meaning of the various dependency types,
i.e., when a dependency type is considered to hold in a certain set of execution traces.
Afterwards, it only rests
to verify that they are computed correctly for every type of commands. This is mostly technical and is not given
in this paper in full detail.

We give the definition for unary dependencies (black and red) first and then the binary ones.

\begin{definition}[unary, black dependencies] \label{def:semantics-blk-unary}
Let $\rho\subset \tseqs$.
We say that $\rho$ satisfies the dependency
$D = \unarydep{i}{\delta}{j}$ 
written $\rho \models D$,
if there are integer constants $d,t,b$, where $d\ge 0$ and $t\ge b\ge 0$, as well as a real constant $c>0$, 
 such that for all 
$\vec x \spge{} t$ there is a sequence
$\sigma = (\vec x, \dots, \vec y) \in \rho$ with $\ecount{\sigma} = b$ satisfying:

\begin{tabular}{llll}
{\upshape (M)} & \multicolumn{3}{l}{$\vec{y} \spge{} \xmin$,} \\
{\upshape (SU1)} & $\delta \ge 1 $ & $\Rightarrow$ & $ y_j \ge x_i$, \\
{\upshape (SU1+)} & $\delta = 1^+ $ & $\Rightarrow$ & $ y_j \ge x_i+\xmin$, \\
{\upshape (SU2)} & $\delta = 2 $ & $\Rightarrow$ & $ y_j \ge 2(x_i - d)$, \\
{\upshape (SU3)} & $\delta = 3 $ & $\Rightarrow$ & $  y_j \ge 2^{c(x_i - d)} $.
\end{tabular}

\noindent
As an exception, 
dependencies $\unarydep{n+1}{1}{n+1}$ are considered to be satisfied by any (non-empty) $\rho$.
\end{definition}


Note that property (M), which also figures in the following definitions, means that no variable has been reset to zero,
for example. Indeed, handling such extensions requires a more difficult analysis~\cite{B2010:DICE}.

Red dependencies express a condition similar to that of black dependencies of the same type,
but they express a dependence on the iteration count
$\ecount{\sigma}$ and they should be satisfied by infinite sets of
traces whose iteration counts form an arithmetic sequence.

\begin{definition}[unary, red dependencies] \label{def:semantics-red-unary}
Let $\rho\subset \tseqs$.
We say that $\rho$ satisfies the red dependency
$D = \unarydep{n+1}{\delta}{j}$ 
written $\rho \models D$,
if there are integer constants $d,t,b,p$, where $p>0$, $d\ge 0$, $t\ge b\ge 0$, 
as well as a real constant $c>0$,
such that for all $i\ge 0$, for all 
$\vec x \spge{} t + ip$,   there is a sequence
$\sigma = (\vec x, \dots, \vec y) \in \rho$, with $\ecount{\sigma} = b + ip$, satisfying:

\begin{tabular}{llll}
{\upshape (M)} & \multicolumn{3}{l}{$\vec{y} \spge{} \xmin$,} \\
{\upshape (SU2)} & $\delta = 2 $ & $\Rightarrow$ & $ y_j \ge 2(\ecount{\sigma}- d)$, \\
{\upshape (SU3)} & $\delta = 3 $ & $\Rightarrow$ & $  y_j \ge 2^{c(\ecount{\sigma} - d)} \,.$
\end{tabular}
\end{definition}

\begin{definition}[binary dependencies] \label{def:semantics-binary}
Let $\rho\subset \tseqs$.
We say that $\rho$ satisfies the dependency $D = \binarydep{i}{j}{k}{\ell}$
written $\rho \models D$,
if there are constants $t\ge b\ge 0$ 
 such that
   for all $\vec x \spge{} t$ there is 
$\sigma = (\vec x, \dots, \vec y) \in \rho$ where $\ecount{\sigma} = b$, satisfying:

\begin{tabular}{ll}
{\upshape (M)} & $\vec{y} \spge{} \xmin$ , \\
{\upshape (SB1)} & $ y_k \ge x_i  \land y_{\ell} \ge x_j  $, \\
{\upshape (SB2)} & $k=\ell \Rightarrow  y_k \ge x_i+x_j$.
\end{tabular}
\end{definition}



Using these definitions we can state the soundness theorem:

\begin{theorem}[lower-bound soundness] \label{thm-soundpol}
If $D\in \depsempar{E}$ then  $\tssempar{E}\models D$.
\end{theorem}

The next subsections justify the soundness for each \lare constructor in turn; this yields Theorem~\ref{thm-soundpol} by simple
induction.

\subsection{Atomic programs}
\label{sec:lin-sound-assign}

This is a trivial part. All atomic programs 
in our core instruction set 
induce obvious dependencies. Note that for a multiplication instruction,
$\X_i \pass \X_j\verb/*X/_k$, we need a
threshold $t \ge 2$ to justify the lower bounds $y_i \ge 2x_j$ and $y_i \ge 2x_k$.

\subsection{Alternation and composition}

Alternation is interpreted as set union in the concrete semantics, and since the property $\rho\models D$ is existential,
we immediately have

\blem
If $\tssempar{E_1} \models D$, or $\tssempar{E_2} \models D$, then
$\tssempar{E_1|E_2} \models D$. 
\elem

For composition, we claim:

\begin{lemma} \label{lem:seq}
If $\tssempar{E_1} \models D_1$ and $\tssempar{E_2}\models D_2$ then
$\tssempar{E_1 E_2}\models D_1\cdot D_2$.
\end{lemma}

This follows by considering $\rho_1,\rho_2 \subset \tseqs$,
such that $\rho_1\models  D_1$ and $\rho_2\models D_2$,
and proving that
$(\rho_1\concat\rho_2) \models D_1\cdot D_2$.
The proof is a tedious case analysis, according to the types of $D_1$ and $D_2$, and follows the
corresponding case in the definition of the product (Definition~\ref{def:prod}).  We will skip it mostly,
giving one case for example, involving binary dependencies:
Suppose that
$D_1 = \binarydep{i}{i'}{j}{j'}$ and $D_2 = \binarydep{j}{j'}{k}{k}$, and $i\ne i'$.
Then $D_1\cdot D_2 = \binarydep{i}{i'}{k}{k}$.
Let $t_1,b_1$ (respectively, $t_2,b_2$) be the constants involved in the application of
definition~\ref{def:semantics-binary} to these dependencies.
Thus there is, for all $\vec{x}\spge{} t_1$ a sequence
$\sigma_1\in \rho_1$ satisfying requirements (M) and (SB1), starting with
$\vec{x}$ and ending with $\vec{y} \spge{} \xmin$. We can restrict
attention to $\vec{x} \spge{} \max(t_1,t_2)$ which implies $\vec{y} \spge{} t_2$. For such $\vec y$ there
will be a $\sigma_2\in \rho_2$, 
 satisfying  $\vec{z} \spge{} \ymin$, plus (M)--(SB2).
 Then
$y_j \ge x_i$, $y_{j'}\ge x_{i'}$, $z_k \ge y_j +  y_{j'}$. 
It follows that $z_k \ge x_i + x_{i'}$,
so the conclusion $(\rho_1\concat \rho_2) \models  D_1\cdot D_2$ holds.

\subsection{Analyzing loops}

 In this analysis we 
assume for simplicity (and omitting the justification) that our experssions are rewritten so that for every 
starred expression there is just one $\ensuresymbol$, in its beginning, thus:
$(\ensuresymbol E)^*$.

Now, recall that $\depsempar{E^*} = LC_\ell(F)  \cdot F$, with $F = LFP(\depsempar{E})$.
We first note that 
if for some finite $m$, $\tssempar{E} \models D_i$ for $i=1,\dots,m$,
then
$\tssempar{(\ensuresymbol E)^*} \models D_1\cdot D_2\ldots\cdot D_m$.
Consequently $\tssempar{(\ensuresymbol E)^*} \models D$ for all $D\in LFP(\depsempar{E})$.
It rests to justify the use of the loop correction operator.

\medskip  

\begin{lemma} 
Let $F = LFP(\depsempar{E})$,  $D\in F$ and $R \in LC_{n+1}(D)$.
Then $\tssempar{(\ensuresymbol E)^*} \models R$.
\end{lemma}

\kern -2ex  

\bprf
Note that if $LC_{n+1}(D) = \emptyset$, there is nothing to prove.
Otherwise, $LC_{n+1}(D) = \{\unarydep{n+1}{\lambda}{i}\}$ for some  $\lambda\ge 2$.
From  $F = LFP(\depsempar{E})$, it clearly follows that, for some $m > 0$,
$D$ is in $(\depsempar{E})^m$; hence 
$(\tssempar{\ensuresymbol E})^{m} \models D$.
%
Denote, for brevity, $\tssempar{\ensuresymbol E}$ by $\rho$.
Checking the cases in the definition of $LC$ (Page~\pageref{def:LC}) we see that $D$ must be $\unarydep{i}{\delta}{i}$ with 
$i\ne n+1$ and $\delta \in \{1^+,2\}$. In particular, $D$ is black.
Hence, there are $t,b,d$ such that
\begin{quote}
   For all $\vec x \spge{} t$ there is a 
$\sigma = (\vec x, \dots, \vec y) \in \rho^{m}$ where $\ecount{\sigma} = b$,
$\vec{y} \spge{} \xmin$ and
$y_i \ge x_i+\xmin$ (for $\delta=1^+$) or $y_i \ge 2(x_i - d)$ (for $\delta=2$).
\end{quote}
Note that $b>0$, since an empty trace only satisfies dependencies of type 1 (for the same reason, $m>0$).

By easy induction we can now derive for any $s>0$, and any $\vec x$ as above, a trace $\pi_s \in \rho^{m s}$
starting at $\vec x$ as above,
and ending with a state $\vec z$ satisfying $\zmin\ge \xmin$ and $z_i\ge x_i+s\,\xmin$ (for $\delta=1^+$)
or $z_i\ge 2^{s} (x_i - d)$ (for $\delta=2$). 
Moreover, $\ecount{\pi_s} = sb$.
Thus,
\begin{align*}
z_i &\ge x_i+s\xmin \ge x_i + (\ecount{\pi_s} / b)\xmin  && \text{(for $\delta=1^+$), or} \\
z_i &\ge 2 ^ {s} (x_i - d) \ge  2^{\ecount{\pi_s} } (x_i - d)     && \text{(for $\delta=2$)}  \,.
\end{align*}
Choosing $c' = 1/b$ we may conclude that if $\ecount{\pi_s}$ is large enough,
and $x_{min} > \max(d,2b,t)$,
we have
\begin{align*}
z_i &\ge 2\cdot \ecount{\pi_s}   && \text{(for $\delta=1^+$), or} \\
z_i &\ge 2^{c'\cdot \ecount{\pi_s}} && \text{(for $\delta=2$)}  \,.
\end{align*}

By the definition of the semantics of the iteration construct,
$\pi_s\in \tssempar{(\ensuresymbol E)^*}$.
The sequences $\pi_s$ thus satisfy the requirements per Definition~\ref{def:semantics-red-unary}, with appropriate
constants (which are not hard to derive from the above discussion, in particular we note that the period of the arithmetic
sequence is $b$).
\eprf

We conclude with the bracket construct. Recall that
\[
 \tssempar{ \,[_\ell E ]\, } = \{ \erase{\sigma} \mid \sigma\in \tssempar{E}, \, \ecount{\sigma}\le (\sigma[0])_\ell \} \,.
\]
Clearly, $\erase{\sigma}$ has the same initial and final states as $\sigma$. Thus a lower bound on the final state of $\sigma$
is valid for $\erase{\sigma}$. Suppose that $\tssempar{E}\models D$; in most cases this implies 
 $\tssempar{ \,[_\ell E ]\, } \models D$ trivially, with the exception being $D = \unarydep{n+1}{\alpha}{k}$, a red dependency.
 In this case the lower bound involves $\ecount{\sigma}$, which ranges over a set $B= \{b, b+p, b+2p, b+3p, \dots\}$. Choosing
 the longest among these sequences that satisfy $\ecount{\sigma}\le (\sigma[0])_\ell $, we obtain 
 $\ecount{\sigma}\ge (\sigma[0])_\ell - p$. Hence, denoting the initial and final states of $\sigma$ by $\vec x$ and $\vec y$,
 respectively, we obtain a result of the form
\begin{align*}
y_i &\ge  2 (x_\ell - p - d)   && \text{(for $\alpha=2$), or} \\
y_i &\ge 2^{c (x_\ell - p - d)}     && \text{(for $\alpha=3$)} 
\end{align*}
which satisfies (SU2) or (SU3), respectively. We also note that (SU1) will be satisfied once the threshold $t$ large enough.

We have now proved the soundness of all analysis rules, and Theorem~\ref{thm-soundpol}  immediately follows.

\section{Upper-Bound Soundness}
\label{sec:completeness}


The upper-bound soundness result will show that the absence of a type-3 dependency for a result variable
implies that it is polynomially bounded in all executions. Thus, here we need an interpretation of the abstract value
(the dependency set) which is \emph{universal} in terms of applying to all traces, and moreover, it has to take into
account \emph{all variables} simultaneously. For intuition, consider an assignment $\X_i \pass \X_j\verb/+X/_k$.
To prove that the result is exponential, it suffices to know that one of $\X_j$, $\X_k$ is exponential. But to prove that
the result is polynomial, we need to know that both of them are.

\paragraph{Multi-polynomials.} 
We introduce the notation $\vec{p}$ for a collection of polynomials $p_{j}$. The
range of the indices is implicit, and should be assumed by the reader to be $\{1,\dots,n\}$ unless the context
dictates otherwise. We may refer to such a collection as a \emph{multi-polynomial}.
Its purpose it to express simultaneous polynomial bounds on several variables.

We say that variable $x_i$ \emph{participates} in polynomial $p(\vec x)$ (or that the polynomial depends on $x_i$)
 if $x_i$ appears in a monomial of $p$ that has a non-zero coefficient.
We use the notation $p[x_k | \pi(k)]$ to indicate that $p$ 
depends only on variables $x_k$
where $k$ satisfies the predicate $\pi$. For example, $p[x_k|k=1]$ 
indicates a polynomial dependent only on $x_1$.

Multi-polynomials can be composed, written $\vec p \circ \vec q$, provided that for every $x_i$ that participates in $\vec p$,
the polynomial $q_i$ is defined.

In the proof we use polynomials
of $n+1$ variables, where $x_1,\dots,x_n$ represent the initial state of the computation under consideration
 while the last variable, $x_{n+1}$, is used
to represent the number of $\ensuresymbol$ symbols in a trace.
Once polynomial bounds of this form (to be called \emph{extended polynomials}) are established for a program $E$,
we obtain bounds for  $[_\ell E]$ by substituting the value
of the loop control variable ${\tt x}_{\ell}$ for $x_{n+1}$.

\paragraph{Dependency matrices.} 
Just as we aggregate upper bounds in a multi-polynomial, we have to aggregate dependencies as well. We use an $(n+1)\times (n+1)$ matrix
$A$ to denote a collection of unary dependencies, i.e., $A_{ij}$ shows the dependence of $x_j$ on $x_i$. Thus, $A\in \matrices$,
where $\deptypes_0 =  \{0,1,1^+,2,3\}$ is the set of dependency types together with 0, a bottom element representing no dependence.
Our matrices have to satisfy a certain validity condition in order to make sense as a set of dependencies. We call
 $A$ \emph{admissible}
 if the following conditions are satisfied:
\begin{align}
& A_{(n+1)(n+1)} = 1
\\
\label{eq:adm}
(\forall i,j) & A_{ij}=1 \Rightarrow A_{kj}=0 \mbox{\rm\ for all } k\ne i  
\,,
\end{align}
where the first condition arises from the special role of $x_{n+1}$, and the second one from the purpose of a dependency
$\unarydep{i}{1}{j}$, which is supposed to mean that $x_j$ obtains its value from $x_i$, without any additions (note the difference
of $1$ to $1^+$).

We denote by $A\le B$ the natural component-wise comparison.

When $S$ and $T$ are sets of matrices, $S\le T$ means
$\forall A\in S \; \exists B\in T \; A\le B$.  

We introduce a ``sum" operation on $\deptypes_0$, denoted by $+$, as follows:
\( 1^+  +  1^+ = 2 \); and
\( \alpha +\beta = \alpha\sqcup\beta\) for all other $\alpha,\beta $.
We then define matrix product in $\matrices$ in the usual way, using the operations $\cdot$ and $+$. 

\paragraph{The set-of-matrices abstraction.} The core idea of this proof is to replace our abstract domain. Instead of sets
of unary and binary dependencies, we generate matrices. The unary dependencies are represented as entries in the matrix,
while binary dependencies correspond to  the co-existence of two 1/$1^+$ entries in the same matrix. However, we do not define
a new abstract interpreter, but obtain the matrices from the sets of dependencies, as follows.

\begin{definition} \label{def-adm-C}
 Let $S$ be a set of dependence facts.
Then $\widehat\somsym(S)$ is the set of all admissible $A$ satisfying
$$
\begin{array}{cl}
\mbox{(M1)} & \forall i,j\ . \ A_{ij}\ne 0 \Rightarrow \longunarydep{i}{A_{ij}}{j} \in S
\\
\mbox{(M2)} & \forall i,j,k,l\ . \ A_{ik},A_{jl}\simeq 1
                        \land (i\ne j \lor k\ne l) \ \Rightarrow\ \binarydep{i}{j}{k}{l} \in S \,.
\end{array}
$$
Let $\somsym(S)$ be the set of maxima of $\widehat\somsym(S)$.
\end{definition}

The \emph{matrix abstraction} of the analysis results for $E$ is the set $\somsempol{E}$
(it seems neat to omit the parentheses in $\somsym(\depsempar{E})$\ ).

\bex
Consider an \lare program $E$ representing the command%
\begin{Verbatim}[codes={\catcode`$=3\catcode`_=8}]
   choose { X$_2$ := X$_3$; X$_3$ := X$_1$+X$_1$} or skip
\end{Verbatim} 
The dependency set for this command is
\[
 \unarydep{1}{1}{1}, \quad
 \unarydep{1}{2}{3},\quad
 \unarydep{2}{1}{2}, \quad
 \unarydep{3}{1}{2}, \quad
 \unarydep{3}{1}{3}, \quad
 \unarydep{4}{1}{4}, \quad
 \binarydep{1}{3}{1}{2}, \quad
 \binarydep{4}{3}{4}{2},
\]
along with all binary dependencies of the form $\binarydep{i}{j}{i}{j}$ (since they all occur in the \texttt{skip} branch).
Here are two elements of $\somsempol{E}$
\[
\bordermatrix{ ~ & {}_1 & {}_2 & {}_3 & {}_4\cr
                          {\scriptstyle 1} & 1 & 0 & 2 & 0\cr
                          {\scriptstyle 2} & 0 & 0 & 0 & 0\cr
                          {\scriptstyle 3} & 0 & 1 & 0 & 0 \cr
                          {\scriptstyle 4} & 0 & 0 & 0 & 1\cr }
\hspace{3em}
\bordermatrix{ ~ & {}_1 & {}_2 & {}_3 & {}_4\cr
                          {\scriptstyle 1} & 1 & 0 & 0 & 0\cr
                          {\scriptstyle 2} & 0 & 1 & 0 & 0\cr
                          {\scriptstyle 3} & 0 & 0 & 1 & 0\cr
                          {\scriptstyle 4} & 0 & 0 & 0 & 1\cr }
\]
Clearly, the first represents the dependencies arising from choosing the first branch, while the second represents the
second branch. Note that a matrix including both $\unarydep{3}{1}{2}$ and $\unarydep{3}{1}{3}$ is excluded
by the constraint (M2). 
\eex

The following lemma (which we cite here without proof) shows that the matrix sets obtained in this way satisfy some constraints
of the kind we would expect a semantic abstraction to fulfill.

\blem \label{lem:matrixsemantics}
The following facts are true for the matrix representation $\somsempol{E}$.
\begin{enumerate}
\item \label{lem:matrix-atomic}
For an atomic program $E$, $\somsempol{E}$ consists of a single matrix. 
\item \label{lem:matrix-choice}
$\somsempol{E_1|E_2} \ge \somsempol{E_1} \cup \somsempol{E_2}$.
\item \label{lem:matrix-seq}
$\somsempol{E_1 E_2} \ge \somsempol{E_1} \cdot \somsempol{E_2} $.
\item \label{lem:matrix-loop}
\be
 \item
  $\somsempol{\LoopExp} \ge  \{ I_{n+1} \}$. 
  \item
  $\somsempol{\LoopExp} \ge \somsempol{\LoopExp} \cdot \left(\somsempol{\LoopExp}\cup \{I_{n+1}\} \right)$.
\ee
\end{enumerate}
\elem

We proceed to give these matrices a meaning in terms of polynomial upper bounds.  The crux of the proof will be to prove that
the matrices computed for a program are sound when interpreted in this fashion.
For simplicity, throughout the rest of the section we assume that our analysis concludes that all variables are polynomially bounded and
prove the soundness of that. Thus, we assume that no $3$-entries occur in our matrices.

\begin{definition}[concretization of dependence vectors] \label{def:Gamma}
Let $v$ be a vector in $\{0,1,1^+,2\}^{(n+1)}$. We define a set of functions, $\Gamma(v)$, to include all polynomials of form
$$\left( \sum_{i\st v_i\simeq 1} a_ix_i \right)  + P[x_i \mid v_i=2]$$
with $a_i \in \{0,1\}$.
Note that if $v$ is the zero vector, we get the constant function 0.
\edfn

\begin{definition}[polynomial upper bounds] \label{def-describe}
Let $\rho\subset \tseqs$.
We say that $\rho$ admits 
a polynomial $p$ as an upper bound on variable $j$, or that $p$ bounds variable $j$ in $\rho$,
if the following holds: 
\begin{equation} \label{eq:description}
\spacedfive{
 \sigma\in\rho,}{
  \tseq{\vec{x}}{\sigma}{\vec{y}},}{
  {x_{n+1} \ge \ecount{\sigma}}}{
  \Longrightarrow}{
 y_j\le p(\vec{x})}
 \,.
\end{equation}
\noindent We also apply this expression to $j=n+1$ (the dummy variable). Here the requirement is $p(\vec x) = x_{n+1}$.
\end{definition}

\begin{definition}[description by a matrix]
Let $\rho\subset \tseqs$ and $A$ an admissible matrix. Suppose that there is an upper bound $p_{A j}$ for variable $j$
in $\rho$, and $p_{A j}\in \Gamma(A_{\bullet j})$.  We then say that $A$ describes variable $j$ in $\rho$. 

We say that 
an admissible matrix $A$ describes $\rho \subseteq \tseqs$ (via a multi-polynomial $\vec p_A$) 
if $\vec p_{A j}\in \Gamma(A_{\bullet j})$ bounds variable $j$ in
$\rho$ for all $j$.
Concisely, we write: ${\rho} \models A : \vec{p}_A $ or, elliptically, 
${\rho}\models A$.
 
We say that a set $\cal A$ of matrices describes $\rho$
if $\rho$ has a cover $\rho \subseteq \bigcup_{A\in{\cal A}}\, \rho_A$ such that
every matrix $A$ describes its corresponding subset $\rho_A$,
i.e., ${\rho_A} \models A$.
We concisely write this statement as
${\rho} \models \mathcal A $.
\end{definition}

Now we have a concise statement of the main theorem of this section:

\begin{theorem}[soundness: upper bounds] \label{thm-sopsound}
For any \lare program $E$,
$\tssempar{E}\models  \somsempol{E}$.
\end{theorem}

\subsection{The proof}
This theorem is proved by structural
induction on $E$.
The base cases are the atomic programs, for which the statement is straight-forward to verify.
 The case of $E_1|E_2$ is also straight-forward, but the case of $E_1 E_2$ is slightly harder. The argument for this case
 should be clear from the statement of the following two lemmata, whose proofs we omit.
 
\begin{lemma} \label{lem-pol-compose}
Let $A$, $B$ be matrices; $p$ a polynomial, $\vec q$ a multi-polynomial, and assume that:
\begin{compactenum}[(1)]
\item For a certain $j$, $p\in \Gamma(B_{\bullet j})$;
\item For all $k$ such that $x_k$ participates in $p$, $q_k$ is defined and $q_k \in \Gamma(A_{\bullet k})$;
\end{compactenum}
Then
$p \circ \vec q \in \Gamma((AB)_{\bullet j})$.
\end{lemma}

\begin{lemma} \label{lem-composition}
Let $\rho_1,\rho_2\subseteq \tseqs$, 
so that $\rho_1\models A : \vec q$ and $\rho_2\models B : \vec p$.
Then  
$\rho_1\concat \rho_2 \models AB: (\vec p \circ \vec q)$.
\end{lemma}

\smallskip
\noindent
Finally we come to the hardest part, which is to prove the following

\smallskip
\begin{lemma} \label{lem-loop-pol}
If $\tssempar{E}\models \somsempol{E}$, then
$\tssempar{\LoopExp}\models \somsempol{\LoopExp}\,$. 
\end{lemma}
\smallskip

For this proof, we let $\MM = \somsempol{E} \cup \{I_{n+1}\}$, and $\MM^* = \somsempol{E^*}\,$.

As the proof is difficult, we only bring some main ideas. The first is the so-called
\emph{size-relation graph}, SRG. This is a graph which shows all the dependencies at once:  its nodes are
$\{1,\dots,n+1\}$ and there is an arc $i\to j$, labeled $\delta$, if $\delta$ is the highest value of entry $A_{ij}$ in
$\MM^*$, and is non-zero.

The decomposition of this graph into strongly connected components (SCCs), and their topological numebring, plays a crucial role.
We show that intra-components arcs must be labeled with $1$ or $1^+$; intuitively in there had been a 2 there, a variable
would have grown exponentially in the loop. The topological ordering of the components induces a
``tiering" of variables so that non-linear data-flow only flows into higher-numbered SCCs.

The second important idea turns up when we want to compute bounds for variables that belong to a non-singleton
SCC.  It turns out to be necessary to compute bounds not only on the values of those variables but also on  the sums of
sets of variables.
We can illustrate this point using the program
\begin{center}
\begin{minipage}{0.9\textwidth}
\begin{Verbatim}[codes={\catcode`$=3\catcode`_=8}]
loop X$_5$ {
  choose
    choose { X$_3$ := X$_1$; X$_4$ := X$_2$ } or { X$_3$ := X$_2$; X$_4$ := X$_1$ }
   or
      X$_1$ := X$_3$ + X$_4$
}
\end{Verbatim}
\end{minipage}
\end{center}
 To prove that there is no exponential growth in this loop,
it is crucial to use the fact that
each of the first two assignments command makes the sum 
$\X_3+\X_4$ equal to $\X_1+\X_2$.  
This implies that subsequently executing the third assignment amounts to setting $\X_1$ to $\X_1+\X_2$; repeatedly
doing so yields polynomial growth.
If we only state bounds on each variable separately we cannot complete
the proof.

Basically, the sets of interest here are sets of variables that exchange their values, but do not duplicate them.
We introduce another definition:

\begin{definition}
Let $C$ be a strongly connected component of the SRG.
We define ${\cal F}(C)$ to be the family of sets
$X\subseteq C$, such that 
for any matrix $A\in \MM^*$, and $i\in C$,
\[
A_{ij}>0\;\land\;A_{ik}>0 \;\land\; j,k\in X\,\Rightarrow\, j=k 
\,. \]  Such a set is called \emph{duplication-free}.
\end{definition}

\bex 
Consider the program shown above and its SRG, shown in Figure~\ref{fig:SRG}.
Let $C$ be the SCC consisting of the nodes
\(
 x_1,\ x_3\ \text{and}\ x_4 \,.
\)
The set $\{1,3,4\}$, representing all three nodes of the component, is \emph{not} duplication-free. In fact, analysis
of the assignment  $\X_1 \pass \X_3+\X_4$ yields dependences
$\unarydep{3}{1^+}{1}$ and $\unarydep{3}{1}{3}$ (among others) as well as binary dependence
$\binarydep{3}{3}{1}{3}$. Consequently, $\MM$ includes a matrix $A$ with $A_{31}=1^+$ and
$A_{33}=1$. Hence any set containing both $1$ and $3$ is not duplication-free. However, the set $\{3,4\}$ is,
and this is used in proving that there is no exponential growth.
\eex

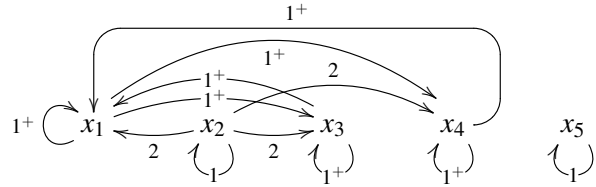
\begin{fig0}{SRG for the example program. Nodes are labeled $x_j$ rather than just $j$ for readability. } {t}{fig:SRG}
\[
\xymatrix@u@R=30pt{
x_1 \ar@(dl,ul)^{1^+}[]\ar@/^10pt/|{\,1^+}[dd]\ar@<5pt>@/^27pt/_{\,1^+}[ddd] & \\
x_2  \ar@(dr,dl)|{1}[]\ar^{2}@/^4pt/[u]\ar_{2}@/_4pt/[d]\ar^{2}@/^15pt/[dd] & \\
x_3  \ar@(dr,dl)|{1^+}[] \ar|{\,1^+}@<-2.5pt>@/_15pt/[uu] & \\
x_4  \ar@(dr,dl)|{1^+}[]    \ar`r[ur] `[uur] `[uuu]_{1^+}[uuu]  & \\ 
x_5  \ar@(dr,dl)|{1}[] & 
}
\]
\bigskip
\end{fig0}

We now summarize in a nutshell the proof of Lemma~\ref{lem-loop-pol}.
The proof constructs, for every $A\in \MM^*$, and every duplication-free set $X$, a multi-polynomial $\vec{h}_{A,X}$,
and proves that every trace in $\tssempar{\LoopExp}$ is upper-bounded by one of these multi-polynomials
(in the sense of Definition~\ref{def-describe}, extended to comprise bounds on 
sets of variables).
We remark that this provides bounds for individual variables since a singleton set is duplication-free.
These polynomials are constructed by induction on the topological ordering of the SCCs, i.e., at every step
of this induction we construct the polynomials, simultaneously, for all of ${\cal F}(C)$ for a component $C$.
There is no space here to give the construction, not to show its soundness; the interested reader is referred to our technical report.


\end{document}